\documentclass[aps,prl,superscriptaddress,twocolumn]{revtex4}

\usepackage[pdftex]{graphicx}
\usepackage{dcolumn}
\usepackage{bm}
\usepackage{color}
\usepackage{amsmath}
\usepackage{centernot}
\usepackage{nicefrac}
\usepackage{amssymb} 

\newcommand{\ff}[1]{{\boldsymbol #1}}
\newcommand{\ca}[1]{{\cal #1}}
\newcommand{\bi}{\begin{itemize}}
\newcommand{\ei}{\end{itemize}}
\newcommand{\be}{\begin{equation}}
\newcommand{\ee}{\end{equation}}
\newcommand{\ba}{\begin{eqnarray}}
\newcommand{\ea}{\end{eqnarray}}

\newcommand{\ket}[1]{| #1 \rangle}
\newcommand{\bra}[1]{\langle #1 |}
\newcommand{\refeq}[1]{Eq.\ (\ref{eq:#1})}
\newcommand{\labeq}[1]{\label{eq:#1}}

\newcommand{\titlepaper}{Geometrical torque on magnetic moments coupled to a correlated antiferromagnet}

\usepackage[pdftex,colorlinks=true,linkcolor=blue,citecolor=blue,filecolor=blue]{hyperref}

\begin{document} 
   
\title{\titlepaper}

\author{Nicolas Lenzing} 

\affiliation{University of Hamburg, Department of Physics, Notkestraße 9-11, 22607 Hamburg, Germany}

\author{David Kr\"uger} 

\affiliation{University of Hamburg, Department of Physics, Notkestraße 9-11, 22607 Hamburg, Germany}

\author{Michael Potthoff}

\affiliation{University of Hamburg, Department of Physics, Notkestraße 9-11, 22607 Hamburg, Germany}

\affiliation{The Hamburg Centre for Ultrafast Imaging, Luruper Chaussee 149, 22761 Hamburg, Germany}

\begin{abstract}
The geometrical spin torque mediates an indirect interaction of magnetic moments, which are weakly exchange coupled to a system of itinerant electrons. 
It originates from a finite spin-Berry curvature and leads to a non-Hamiltonian magnetic-moment dynamics.
We demonstrate that there is an unprecedentedly strong geometrical spin torque in case of an electron system, where correlations cause antiferromagnetic long-range order. 
The key observation is that the anomalous torque is strongly boosted by low-energy magnon modes emerging in the two-electron spin-excitation spectrum due to spontaneous breaking of SU(2) spin-rotation symmetry. 
As long as single-electron excitations are gapped out, the effect is largely universal, i.e., essentially independent of the details of the electronic structure, but decisively dependent on the lattice dimension and spatial and spin anisotropies.
Analogous to the reasoning that leads to the Mermin-Wagner theorem, there is a lower critical dimension at and below which the spin-Berry curvature diverges.
\end{abstract} 

\maketitle 

\paragraph{\color{blue} Introduction.}  

A magnetic moment coupled to a system of itinerant electrons via a local exchange interaction of strength $J$ experiences a spin torque which leads to precession dynamics. 
For several magnetic moments $\ff S_m$ (with $m=1,...,M$), usually described as classical fixed-length spins, there are further torques caused by, e.g., indirect exchange interactions mediated by the electron system. 
These {\em Hamiltonian} spin torques, well known in micromagnetics \cite{BMS09} and in the theory of coupled spin-electron dynamics  \cite{KFN09,BNF12,EFC+14,SP15,SRP16a,CBW+18,BN19}, all derive from interaction terms in the quantum-classical Hamiltonian \cite{Elz12} for the spin and electron degrees of freedom. 
In addition, there is a non-Hamiltonian spin torque that has a purely {\em geometric} nature.
This geometrical spin torque represents the feedback of the Berry physics \cite{Ber84} on the classical magnetic-moment dynamics.

Generally, such feedback effects have been pointed out early \cite{KI85,MSW86,Zyg87} but have not been studied in spin dynamics theory until recently \cite{SP17}.
For weak $J$ compared to the typical energy scales of the electron system, the classical spin dynamics is slow, such that the electron system accumulates a geometrical phase which is gauge independent in case of a cyclic motion \cite{Ber84,Sim83,WZ84}.
This Berry phase is closely related to the Berry curvature, a two-form which, when integrated in classical parameter space over a two-dimensional surface bounded by a closed path $\ca C$ yields the Berry phase associated with $\ca C$.
For example, in molecular physics \cite{BMK+03} and when treating the coordinates of the nuclei classically, the feedback of the Berry physics produces an additional geometrical force, where the Berry curvature plays the role of a magnetic field in the nuclei equations of motion. 
This effect is known as ``geometrical magnetism'' \cite{BR93b,CDH12}.

The geometrical spin torque resulting from the spin-Berry curvature (SBC) \cite{SP17} is the analogous concept in the field of atomistic spin dynamics \cite{SHNE08,EFC+14}.
As opposed to the closely related geometrical friction term \cite{BR93b,CDH12}, i.e., Gilbert damping \cite{llg}, it is energy conserving. 
But, importantly, the SBC is non-Hamiltonian and emerges for weak $J$, i.e., in the limit of slow classical spin dynamics.
However, the effects are typically weak \cite{MP22} for a solid \cite{Ihm91}, such that it appears difficult to disentangle the effect of the geometrical spin torque from other contributions \cite{BN20}. 

\begin{table}[b]
\caption{
\label{tab:dim} 
Spin-Berry curvature of a spontaneously symmetry-broken antiferromagnetic state with gapped single-particle excitations. 
$\ff k$: wave vector.
See text for discussion.
}
\begin{tabular}{|c|c|c|c|c|}
\hline
\hline
lattice & spin-Berry & distance & magnetic  \\
dimension & curvature (SBC) & dependence& ground state \\
\hline
\hline
1 &  divergent   & --- & ---\\
2 &  log. divergent   & --- & stable\\
3 &  regular   & $1/R$ & stable\\
\hline
\\[-3.5mm]
$D\ge 4$ &  $\sim \int_{0}^{\Lambda_{\sf cutoff}} dk \, k^{D-3}$ & $1/R^{D-2}$ & stable \\[0.5mm]
\hline
\hline
\end{tabular}
\end{table}

In this Letter we study the geometrical spin torque for magnetic moments coupled to a magnetic solid: a correlated $D$-dimensional antiferromagnetic (AF) insulator.
This is a generic situation realized, e.g., by magnetic impurities in the bulk or by magnetic adatoms on the surface of the antiferromagnet.
We demonstrate that the magnitude of the SBC is governed by the magnon-excitation spectrum. 
This has very general consequences:
the SBC must diverge for $D=1$ but is regular for $D \ge 3$, see Tab.\ \ref{tab:dim}.
For $D=2$ the SBC generically exhibits a logarithmic divergence as a function of any perturbation causing a gap in the magnon dispersion, such as magnetic anisotropies or external magnetic fields. 
The magnitude of the SBC and thus the impact on the magnetic-moment dynamics is studied for the Hubbard model at half-filling and zero temperature as a prototype of a correlation-induced insulator.

\paragraph{\color{blue} Time-reversal symmetry (TRS).}  
Within adiabatic spin-dynamics theory \cite{SP17,MP22}, the geometrical spin torque is obtained from the SBC of the electron system, see \refeq{geo} below. 
Importantly, a finite SBC generally requires TRS breaking in the electron system \cite{MP22}.
If $J$ is strong, as assumed in Ref.\ \onlinecite{SP17}, TRS is broken by the classical spin moment itself, as this acts like a local symmetry-breaking field. 
TRS breaking can be waived only at the cost of working with a non-Abelian extension of the theory well beyond the adiabatic limit \cite{LLP22}, where the dynamics is governed by the generically finite non-Abelian spin-Berry curvature.
Another approach is to replace the electron system by an entirely classical model composed of ``slow'' and ``fast'' spin moments \cite{EMP20,MP21}.
This circumvents the necessity of TRS breaking altogether but still exhibits the feedback of holonomy effects in purely classical systems \cite{Han85}.
For magnetic moments coupled to {\em quantum} systems and in the physically relevant weak-$J$ regime, a finite SBC can be achieved with an external magnetic field, or with a (staggered) orbital field as considered recently \cite{MP22} with the Haldane model \cite{Hal88} as a prototype of a TRS-breaking Chern insulator \cite{Ber13}.
However, fine tuning of the parameters is required to achieve considerable effects \cite{MP22}.
Here we consider an electron system in which correlations induce a TRS-breaking AF state.
The AF order not only enables a finite SBC but also strongly boosts its magnitude due to magnon modes in the spin-excitation spectrum.

\paragraph{\color{blue} Dynamics of magnetic moments.}  

We are interested in the slow dynamics of $M$ magnetic moments, described as classical spins $\ff S_{m}$ of unit length, which are coupled to a correlated electron system with Hamiltonian $H_{\rm el}$ via a local exchange interaction 
$H_{\rm int} = J \sum_{m=1}^{M} \ff s_{i_{m}} \ff S_{m}$. 
Here, $i_{m}$ is the site, at which the $m$-th moment is coupled to, and $\ff s_{i} = 1/2 \sum_{\sigma\sigma'} c^{\dagger}_{i\sigma} \ff \tau_{\sigma\sigma'} c_{i\sigma'}$, where $\ff \tau$ is the vector of Pauli matrices, is the local spin moment at site $i$ of the electron system.
The total Hamiltonian is $H=H(\ff S) = H_{\rm el} + H_{\rm int}(\ff S)$ and depends on the configuration $\ff S = (\ff S_{1}, ..., \ff S_{m})$ of the magnetic moments.

Assuming that the electron system at any instant of time $t$ is in its instantaneous ground state for the spin configuration $\ff S(t)$, i.e., $|\Psi(t) \rangle = |\Psi_{0}(\ff S(t))\rangle$, the equation of motion of adiabatic spin dynamics is given by \cite{SP17,MP22}
\be
\dot{\ff S}_{m} 
= 
(\ff T_{m}^{(\rm H)} + \ff T_{m}^{\rm (geo)}) \times \ff S_{m} 
\: .
\labeq{eom}
\ee
Here $\ff T_{m}^{\rm (H)} \times \ff S_{m}$ with $\ff T_{m}^{\rm (H)} = \partial \langle H(\ff S) \rangle / \partial \ff S_{m} = J \langle \ff s_{i_{m}} \rangle$ is the conventional (Hamiltonian) spin torque, where $\langle \cdots \rangle$ is the instantaneous ground-state expectation value.

\paragraph{\color{blue} Geometrical spin torque.}  

The second term, the geometrical spin torque $\ff T_{m}^{\rm (geo)} \times \ff S_{m}$ is necessary to enforce the constraint $|\Psi(t) \rangle = |\Psi_{0}(\ff S(t))\rangle$ and has been derived within a quantum-classical Lagrange formalism in Refs.\ \onlinecite{SP17,MP22}.
This assumes that the ground state is non-degenerate (otherwise non-Abelian spin-dynamics theory \cite{LLP22} must be used) and that $J$ is sufficiently weak so that the classical spin dynamics is much slower than typical relaxation time scales of the quantum system $H_{\rm el}$.
Alternatively, the term may be derived within adiabatic response theory \cite{BR92,BR93b,CDH12} as the first nontrivial correction in a systematic expansion of the response of a driven system with respect to the driving speed, when applied to spin dynamics \cite{LP23}.
It is given by
\be
\ff T_{m}^{\rm (geo)}
=
\sum_{\alpha} \sum_{m'\alpha'} \Omega_{m'm,\alpha'\alpha}(\ff S) \dot{S}_{m'\alpha'} \ff e_{\alpha} 
\; ,
\labeq{geo}
\ee
with $\alpha = x,y,z$ and the $\alpha$-th unit vector $\ff e_{\alpha}$, and where
\be
\Omega_{mm',\alpha\alpha'} (\ff S )
=
\frac{\partial}{\partial S_{m\alpha}} A_{m'\alpha'}(\ff S )
-
\frac{\partial}{\partial S_{m'\alpha'}} A_{m\alpha}(\ff S )
\labeq{curv}
\ee
is the spin-Berry curvature. At each spin configuration $\ff S$, this is a real antisymmetric tensor
($\Omega_{m'm,\alpha'\alpha}=-\Omega_{mm',\alpha\alpha'}$), which is invariant under local gauge transformations of the ground states $\ket{\Psi_{0}(\ff S)} \mapsto e^{i \phi(\ff S)} \ket{\Psi_{0}(\ff S)}$. 
It is the exterior derivative of the spin-Berry connection $\ff A_{m} = i \langle \Psi_{0} | \frac{\partial}{\partial \ff S_{m}} | \Psi_{0} \rangle$, which describes parallel transport of the ground state $\ket{\Psi_{0}(\ff S)}$ on the manifold of spin configurations $\ca M$.
For $M$ classical spins $\ff S_{m} \in S^{2}$, this is given by the $M$-fold Cartesian product of 2-spheres $\ca M \equiv S^{2} \times \cdots \times S^{2}$.

\paragraph{\color{blue} Spontaneous antiferromagnetic order.}  

We consider a coupling of the magnetic spin moments to the single-band Hubbard model \cite{Geb97,EFG+05} on a $D$-dimensional hypercubic lattice as a prototypical model for itinerant magnetic order. 
Its Hamiltonian is $H_{\rm el} = -t\sum_{ij}^{\rm {n.n.}} \sum_{\sigma=\uparrow, \downarrow} c^{\dagger}_{i\sigma} c_{j\sigma} + U \sum_{i} n_{i\uparrow} n_{i\downarrow}$, where the nearest-neighbor hopping $t=1$ fixes the energy and (with $\hbar \equiv 1$) the time scales.
$c_{i\sigma}$ annihilates an electron at site $i$ with spin projection $\sigma$, and $n_{i\sigma} = c_{i\sigma}^{\dagger} c_{i\sigma}$. 
The sums over $i, j$ are restricted to nearest neighbors, and $L$ is the total number of sites. 
It is well known \cite{And50,SWZ89,Man91,Aue94,SGR+15}
that at half-filling, repulsive Hubbard-$U$ and for $D \ge 2$, the ground state of the system in the thermodynamical limit $L\to \infty$ develops long-range AF correlations. 
SU(2) spin-rotation symmetry and therewith TRS are spontaneously broken, and the ordered state is characterized by a finite staggered magnetization $\ff m=m\ff e_{z}$ with $m = L^{-1} \sum_{i} z_{i} \langle n_{i\uparrow} - n_{i\downarrow} \rangle$ and $z_{i}=\pm 1$ for $i$ in sublattice A or B, respectively.
We assume $m>0$ for sublattice A.

At weak $U$, AF order is driven by the Slater mechanism and perturbatively accessible \cite{SWZ89,Aue94}.
Within self-consistent Hartree-Fock theory \cite{Mor85}, the one-electron excitation spectrum displays a gap $\Delta = U m$ at wave vector $Q=(\pi,\pi, ...)$ in the conventional Brillouin zone.
The two-electron spin-excitation spectrum is well described by standard random-phase approximation (RPA) but for the symmetry-broken AF state \cite{RKEH12,DT21,LW60,RHT+18}. 

In the strong-$U$ limit, the one-electron spectrum is dominated by a large Hubbard gap $\Delta \sim U$ and well developed local spin moments, coupled via Anderson's superexchange \cite{And50,Aue94}. 
Here, the model maps onto the Heisenberg spin-1/2 Hamiltonian with AF exchange $J_{\rm H} = 4t^{2}/U$ and AF long-range order, see Refs.\ \onlinecite{SHN89,Man91,San97}, for example.
To compute the low-energy magnon dispersion and states, we can apply spin-wave theory (SWT) \cite{And52} to the AF Heisenberg model and use the Holstein-Primakoff transformation \cite{HP40} at linear order. 
Linear SWT is motivated by the fact that single-magnon decay requires overlap with the two-magnon continuum, so that the picture of a stable magnon gas is protected by kinematic restrictions at low energies \cite{HWA92,CM16,MDG+18,McC22}.

\paragraph{\color{blue} Spin-Berry curvature of an antiferromagnet.}  

To compute the geometrical spin torque, we make use of a Lehmann-type representation of the SBC starting from Eq.\ (\ref{eq:curv}). 
This is straightforwardly derived \cite{MP22} using a resolution of the the unity, $\ff 1 = \sum_{n} | \Psi_{n}(\ff S) \rangle \langle \Psi_{n}(\ff S) |$, with an orthonormal basis of instantaneous eigenstates of $H_{\rm el} + H_{\rm int}(\ff S)$:
\be
\Omega_{mm',\alpha\alpha'}
= 
- 2 J^{2}  \mbox{Im} \!
\sum_{n \neq 0}
\frac{\bra{\Psi_{0}} s_{i_{m}}^{ \alpha} \ket{\Psi_{n}} \, \bra{\Psi_{n}} s_{i_{m'}}^{\alpha'} \ket{ \Psi_{0} }}{(E_n - E_0)^2} 
\: .
\labeq{lehmann}
\ee
Note that, due to the $J^{2}$ prefactor, the $\ff S$ dependence of the eigenenergies and eigenstates will provide corrections to \refeq{lehmann} only at order $J^{3}$. 
As we refer to the weak-$J$ limit, these will be neglected in the following.

In the AF phase and assuming that the order parameter is aligned to the $z$ axis, $\langle \ff s_{i} \rangle = (-1)^{i} m \ff e_{z}$, there is a remaining SO(2) symmetry of the energy eigenstates under spin rotations around $\ff e_{z}$. 
This unbroken spin-rotation symmetry, together with the spatial inversion and translation symmetries of $H_{\rm el}$, and the antisymmetry $\Omega_{mm',\alpha\alpha'} = - \Omega_{m'm,\alpha'\alpha}$ [see Eq.\ (\ref{eq:curv})] imply that the spin-Berry curvature tensor is entirely fixed by a single real number $\Omega \equiv \Omega_{mm',xy} = - \Omega_{mm',yx}$ for each fixed pair of sites $i_{m}$, $i_{m'}$. 
All other elements must vanish, as is detailed by the symmetry analysis in Sections A and B of the Supplemental Material (SM) \cite{SM}. 

In a first step, for weak $U$, we compute the SBC via
\be
\Omega_{mm'} 
=
- i J^{2}
\frac{\partial}{\partial \omega} \chi_{i_{m} i_{m'},xy}(\omega) \Big|_{\omega =0}
+
\ca O(J^{3})
\: ,
\labeq{omchi}
\ee
where $\chi_{ii',\alpha\alpha'}(\omega) = L^{-1} \sum_{\ff k} e^{i\ff k (\ff R_{i} - \ff R_{i'})} \chi_{\alpha\alpha'}(\ff k,\omega)$ is the real-space retarded susceptibility, obtained by the RPA (see SM, Sec.\ C \cite{SM}). 
The relation \refeq{omchi} is easily derived by comparing the representation \refeq{lehmann} of the SBC with the Lehmann representation of the susceptibility (SM, Secs.\ A, B \cite{SM}).
Therewith, the susceptibility in the symmetry-broken AF state is seen to play a dual role for the spin dynamics: 
(i) via \refeq{omchi} and \refeq{geo} its frequency derivative at $\omega=0$ yields the geometrical spin torque $\ff T_{m}^{\rm (geo)} \times \ff S_{m}$, and 
(ii) the static susceptibility yields, in the weak-$J$ regime, the conventional RKKY spin torque 
$\ff T_{m}^{\rm (H)} \times \ff S_{m}$ with $\ff T_{m}^{\rm (H)} = \partial H_{\rm RKKY} / \partial \ff S_{m}$, 
where $H_{\rm RKKY} = J^{2} \sum \chi_{i_{m} i_{m'},\alpha \alpha'}(\omega=0) S_{m\alpha} S_{m'\alpha'}$ is the perturbative RKKY Hamiltonian of the AF state.

\begin{figure}[t]
\includegraphics[width=0.99\columnwidth]{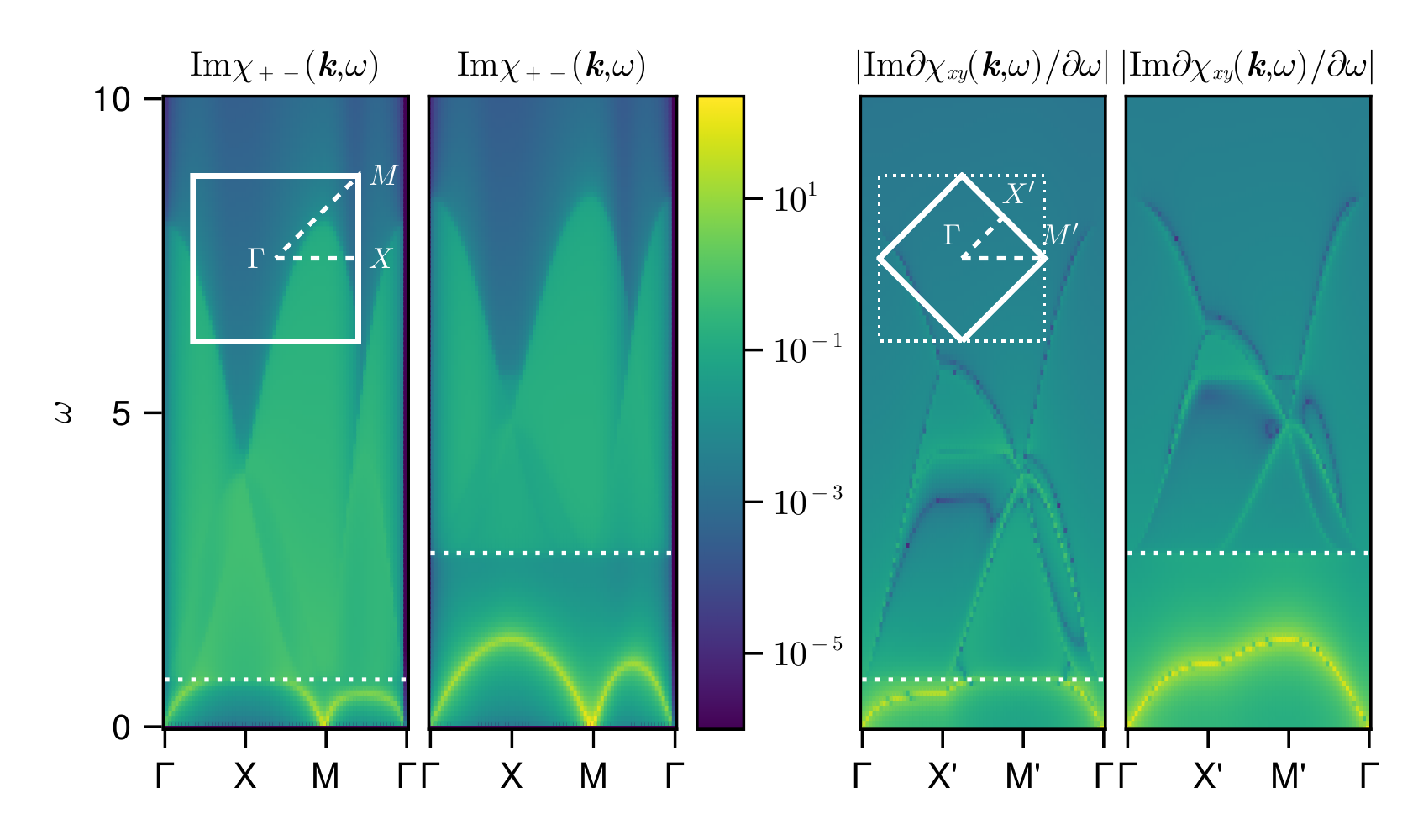}
\caption{
{\em Left:} Transversal retarded ground-state spin susceptibility $\mbox{Im} \, \chi_{+-}(\ff k, \omega)$ for $U=2$ and $U=4$ along high-symmetry directions in the conventional $D=2$ Brillouin zone, as obtained by RPA.
{\em Right:} Frequency derivative $\mbox{Im} \, \partial_\omega \chi_{\rm xy}(\ff k,\omega)$ (absolute values) in the mBz, related to the SBC at $\omega=0$. 
White dotted lines: Slater gap $\Delta=Um$ (onset of the continuum).
Lorentzian broadening $\omega\to \omega+i\eta$ with $\eta=0.045$. 
Energy scale: $t = 1$.
}
\label{fig:chi}
\end{figure}

For the Hubbard model on the $D=2$ square lattice the spin-excitation spectrum $\chi_{+-}(\ff k,\omega)$, see Fig.\ \ref{fig:chi} (left) for $U=2$ and $U=4$, consists of a continuum at high frequencies $\omega > \Delta = U m$ ($\Delta \approx 0.75$ for $U=2$, $\Delta \approx 2.76$ for $U=4$) and, furthermore, within the gap an undamped transversal and doubly degenerate magnon mode. 
This mode takes most of the spectral weight. 
The magnon contribution to the derivative $\partial_\omega \chi_{xy}(\ff k,\omega)$ on sublattice A (Fig.\ \ref{fig:chi}, right) is even more pronounced, especially for $\omega=0$, where it is related to the SBC by \refeq{omchi}.

\paragraph{\color{blue} Goldstone theorem, implications.}  

In our second step, we exploit the fact that the spin-excitation spectrum of an AF insulator has a universal structure at low frequencies.
This is due to Goldstone's theorem which enforces the presence of gapless magnon modes \cite{Gol61,YJ61,Bra10}. 
In the collinear AF state and corresponding to the two broken generators of the spin SU(2) symmetry, there are two degenerate modes with a linear and isotropic dispersion in the vicinity of the $\Gamma$ point in the magnetic  Brillouin zone (mBz). 
Linear SWT applied to the Heisenberg model that emerges in the strong $U$ limit captures this physics, i.e., the dispersion close to $\Gamma$ is given by $\frac12 J_{\rm H} \omega(\ff k) = c_{\rm s} k + \ca O(k^{2})$, where $c_{\rm s}$ is the spin-wave velocity.
Using the magnon energies and eigenstates, we can compute the SBC in this limit from \refeq{lehmann} directly (SM, Secs.\ D, E \cite{SM}), ending up with 
\be
\Omega_{mm^\prime} 
= 
\mp \frac{2J^2}{J_{\rm H}^2} 
\frac{1}{(2\pi)^{D}}
\int_{\rm mBz} \!\! d^{D}k \frac{\cos(\boldsymbol{k}(\ff R_{i_{m}} \! - \! \ff R_{i_{m'}}))}{\omega(\ff k)^2}
\: ,
\labeq{swt}
\ee
if both, $i_{m},i_{m'}$ belong to sublattice A ($-$ sign) or B ($+$ sign), and $\Omega_{mm^\prime} = 0$ else.

For $D=2$, the linear dispersion close to $\Gamma$ then implies a $1/k^{2}$ singularity of the integrand and thus a logarithmic infrared divergence.
For $D\ge 3$, the local ($m=m'$) SBC is finite. 
We note that the same arguments as invoked for the Mermin-Wagner theorem \cite{And52,MW66}, i.e., a divergence due to the low-energy spin excitations, here lead to a lower critical dimension ($D_{\rm c}=3$) that is shifted by one, see Tab.\ \ref{tab:dim}.
The numerical value for the $D=3$ local SBC is $\Omega_{\rm loc} \approx -0.084 \, J^2/J_{\rm H}^2 = -0.084 \,J^2 U^{2}/16 t^{4}$. 
When scaling the hopping as $t = t^{\ast} / \sqrt{D}$ with $t^{\ast}=\mbox{const}$ \cite{MV89,MH89b}, the modulus of the SBC decreases monotonically with $D$, and the SBC approaches a finite mean-field value $|\Omega_{\rm loc}| \to J^{2} U^{2}/32t^{\ast}{}^{4}$ for $D \to \infty$ (SM, Sec.\ F \cite{SM}).

\begin{figure}[t]
\includegraphics[width=0.85\columnwidth]{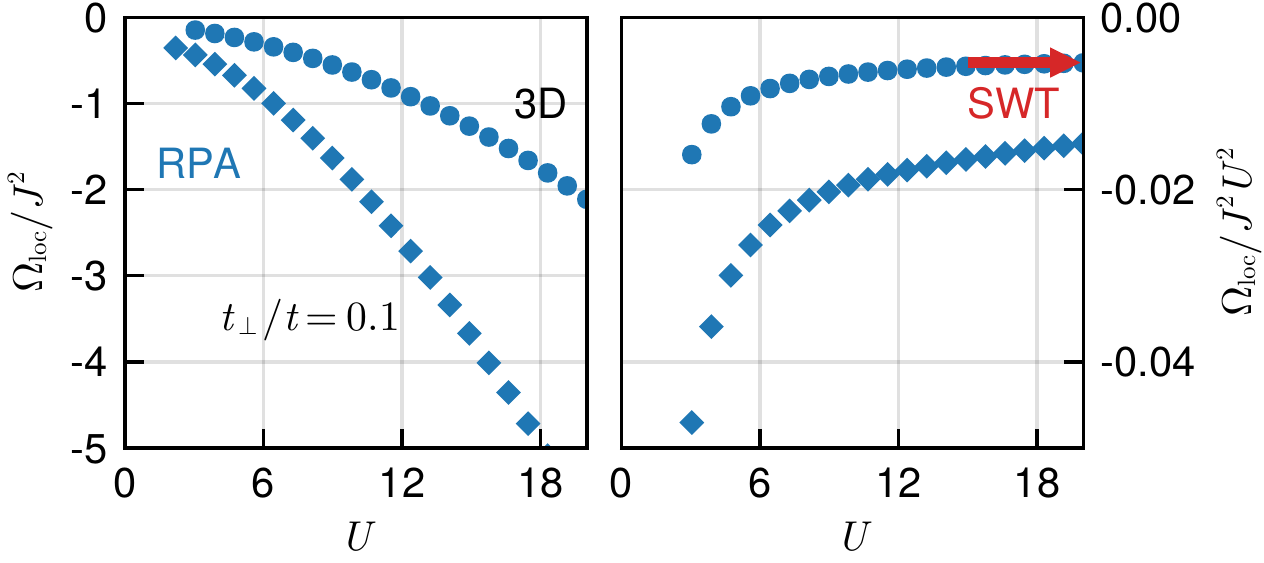}
\caption{
Local SBC as function of $U$ for $D=3$, as obtained from RPA. 
{\it Left:} $\Omega_{\rm loc} / J^{2}$. 
{\it Right:} $\Omega_{\rm loc} / J^{2} U^{2}$. 
Diamonds: $t_\perp / t =0.1$, see also Fig.\ \ref{fig:adep} (left). 
Red arrow: $D=3$ SWT ($U\to \infty$) result. 
$\eta=0.035$.
}
\label{fig:udep}
\end{figure}

\paragraph{\color{blue} Magnitude of the SBC.}  

SWT predicts a $U^{2}$ dependence of the SBC in the Heisenberg limit for strong $U$. 
For $U=0$, on the other hand, TRS of the resulting paramagnetic state implies that it must vanish.
For $U\to 0$, there is an intricate competition between the exponential suppression of the order parameter $m\propto e^{-1/U}$, i.e., of the ``strength'' of TRS breaking and thus of the SBC and, on the other hand, the exponential closure of the single-electron Slater gap $\Delta=Um$ and thus of the onset of the continuum in the spin-excitation spectrum resulting in continuum contributions that favor a large SBC.
Our numerical results for the local SBC in $D=3$, as obtained from weak-coupling RPA and strong-coupling SWT are displayed in Fig.\ \ref{fig:udep}. 
With increasing $U$ we find a smooth crossover from the Slater to the Heisenberg limit with a monotonically increasing $|\Omega_{\rm loc}|$. 

The nonlocal SBC at large distances $R \equiv \| \ff R_{i_{m}} - \ff R_{i_{m'}} \|$ is again governed by the linear dispersion at low frequencies. 
Carrying out the integration in Eq.\ (\ref{eq:swt}) for $R \to \infty$ we find $\Omega(R) \propto 1/R^{D-2}$ (see Tab.\ \ref{tab:dim} and SM, Sec.\ F \cite{SM}).
For $D=3$ this implies that the geometrical spin torque mediates a long-range coupling in the spin dynamics. 

Compared to previous studies \cite{SP17,LLP22,MP22,MP21,BN20,EMP20} the $D=3$ value of the local SBC 
$|\Omega_{\rm loc}| \approx 0.084 \, J^2/J_{\rm H}^2$ 
is several orders of magnitude larger for realistic parameters $J, J_{\rm H} \ll t, U$.
Renomalization of $c_{s} \to c'_{s} \approx 1.1 c_{s}$ due to magnon interaction \cite{Ogu60} leads to a slightly smaller SBC, $|\Omega_{\rm loc}| \to (c_{\rm s}/c_{\rm s}')^{2} |\Omega_{\rm loc}|$.

\begin{figure}[t]
\includegraphics[width=0.95\columnwidth]{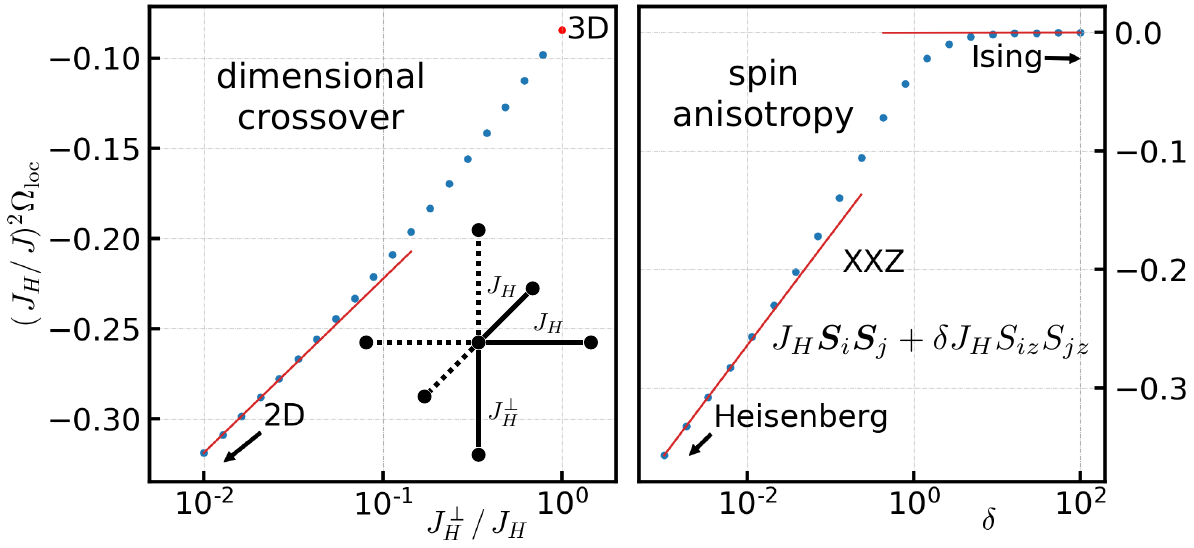}
\caption{
SWT results (dots) for anisotropic systems. 
{\em Left,  dimensional crossover:}
local SBC for $D=3$ but with a spatially 
anisotropic nearest-neighbor Heisenberg exchange $J_{\rm H}^{\perp} \le J_{\rm H}$.
{\em Right, spin anisotropy}:
SBC as function of the coupling anisotropy parameter $\delta$.
}
\label{fig:adep}
\end{figure}

There are at least two routes that lead to an even larger $|\Omega_{\rm loc}|$: 
Namely, we can take advantage of the formally infinite SBC in $D=2$ and regularize the theory
(i) by dimensional crossover to $D=3$ \cite{BG90,MSS92,MSS93}, i.e., by switching on a small hopping $t_{\perp}$ in the third dimension (Fig.\ \ref{fig:udep}) implying $J_{\rm H}^{\perp} \ll J_{\rm H,x} = J_{\rm H,y} = J_{\rm H}$, see Fig.\ \ref{fig:adep} (left), or
(ii) by switching on a magnetic anisotropy to open a small gap in the magnon spectrum (Fig.\ \ref{fig:adep}, right), i.e., by adding an Ising term $\delta \, J_{\rm H} S_{iz} S_{jz}$ to the standard Heisenberg coupling $J_{\rm H} \ff S_{i} \ff S_{j}$.
A moderate $J_{\rm H}^{\perp}/J_{\rm H} = 0.1$ yields a SBC $|\Omega_{\rm loc}| \approx 0.22 J^{2}/J_{\rm H}^{2}$. 
About the same enhancement is obtained for an anisotropy parameter $\delta \sim 10^{-2}$.  

\paragraph{\color{blue} Geometrical spin dynamics.}  

For the AF ordered phase, \refeq{eom} tells us that the dominating effect in the magnetic-moment dynamics is a precession around the staggered magnetization $\ff m$ on a time scale $1/J$. 
This effect dominates the weaker (and slower) anisotropic RKKY-type exchange on the scale $J^{2}$. 
Importantly, the SBC $\Omega \sim J^{2}$ enters the equations of motion as a renormalization {\em factor} (for $M>1$ classical spins as a matrix factor) rather than a summand and thus does {\em not compete} with the stronger direct exchange of order $J$
(SM, Sec.\ G \cite{SM}).
For $M=1$ this factor amounts to $1/(1-\Omega_{\rm loc} S_{z})$, such that the most pronounced effects are found for a SBC of intermediate strength, $\Omega_{\rm loc} = \ca O(1)$. 
This holds true for $M=2$ as well, as is detailed in the SM, Sec.\ G \cite{SM}. 
Note that a singular renormalization indicates a breakdown of the theory as this is the point where the condition for nearly adiabatic spin dynamics is invalidated. 
Note further that the precession comes with an inverted orientation beyond the singular point. 

\paragraph{\color{blue} Conclusions.}  

A hitherto unknown but generic interplay of electron correlations, spontaneous symmetry breaking, gapless Goldstone bosons, and a holonomy on the configuration space of classical spin degrees of freedom leads to non-Hamiltonian effects, such as renormalization of precession frequencies, inverted orientation of the precessional motion, or long-range interactions, in the spin dynamics.
This is due to a geometrical spin torque which is finite for correlated AF ground states in lattice models with dimension $D \ge 3$ and diverges for $D \le 2$, caused by the same mechanism that leads to the Mermin-Wagner theorem, however, shifted by one dimension.
With a SBC $\Omega_{\rm loc} = \ca O(1)$ for typical parameters, the effect is unexpectedly large. 
It is boosted by electron correlations and further enhanced by spatial and by spin anisotropies. 
We expect a strong overall impact on the phenomenology of atomistic spin dynamics.

\paragraph{\color{blue} Acknowledgments.}  
This work was supported by the Deutsche Forschungsgemeinschaft (DFG, German Research Foundation) through the research unit QUAST, FOR 5249 (project P8), project ID 449872909, and through the Cluster of Excellence ``Advanced Imaging of Matter'' - EXC 2056 - project ID 390715994. 



\begin{widetext}
\newpage
\mbox{}
\setcounter{page}{1}

\begin{center}
{\large \bfseries 
\titlepaper

\mbox{}\\

--- Supplemental Material ---
}

\mbox{}\\

{
Nicolas Lenzing,$^{1}$
David Kr\"uger,$^{1}$
and
Michael Potthoff$^{1,2}$
}

\mbox{}\\[-2mm]

{
\small \it 
$^{1}$
University of Hamburg, Department of Physics, Notkestraße 9-11, 22607 Hamburg, Germany

$^{2}$The Hamburg Centre for Ultrafast Imaging, Luruper Chaussee 149, 22761 Hamburg, Germany
}

\end{center}

\end{widetext}

\paragraph{\color{blue} 
Section A: SO(2) symmetry analysis.} 

The retarded spin susceptibility is defined as 
\be
  \chi_{ii',\alpha\alpha'}(t) = - i \Theta(t) \langle [ s_{i\alpha} (t), s_{i'\alpha'}(0)] \rangle \: ,
\labeq{chi1}
\ee
where $\Theta$ is the step function, $s_{i\alpha} (t) = e^{iH_{\rm el}t} s_{i\alpha} e^{-iH_{\rm el}t}$, and $\langle \cdots \rangle$ is the ground-state expectation value.
Fourier transformation to frequency space yields the Lehmann representation in terms of an energy eigenbasis $\{|\Psi_{n}\rangle\}$:
\ba
  \chi_{ii',\alpha\alpha'}(\omega) 
  & = &
  \sum_{n} 
  \Bigg( \frac{\langle \Psi_{0} | s_{i\alpha} | \Psi_{n} \rangle \langle \Psi_{n} | s_{i' \alpha'} | \Psi_{0} \rangle}{\omega + i \eta - (E_{n} - E_{0})}
\nonumber \\
  & - &
  \frac{\langle \Psi_{0} | s_{i'\alpha'} | \Psi_{n} \rangle \langle \Psi_{n} | s_{i \alpha} | \Psi_{0} \rangle}{\omega + i \eta - (E_{0} - E_{n})}
  \Bigg)
\: .
\labeq{chileh}
\ea
We have the relation $\chi_{ii',\alpha\alpha'}(\omega)^{\ast} = \chi_{ii',\alpha\alpha'}(-\omega)$. 
The spectral density $-(1/\pi) \mbox{Im} \chi_{ii',\alpha\alpha'}(\omega) = \chi_{ii',\alpha\alpha'}(\omega) - \chi_{ii',\alpha\alpha'}(-\omega) / 2i$ is an antisymmetric function of $\omega$.

In the AF phase with order parameter $\ff m=m \ff e_{z}$, there is a remaining SO(2) symmetry of the nondegenerate energy eigenstates under spin rotations around $\ff e_{z}$, which is unitarily represented by $U_{\rm R}=e^{-i s_{\rm tot, z} \varphi}$ on the Fock space with the $z$-component of the total spin $\ff s_{\rm tot} = \sum_{i} \ff s_{i}$ as the unbroken generator and the rotation angle $\varphi$.
We have $U_{\rm R}^{\dagger} | \Psi_{n} \rangle = e^{i\phi_{n}} | \Psi_{n} \rangle$ with phases $\phi_{n}$. 
Since $\ff s_{i}$ is a vector operator, we have $U_{\rm R}^\dagger s_{i\alpha} U_{\rm R} = \sum_{\beta} R_{\alpha\beta} s_{i\beta}$, where $\underline{R}=\underline{R}(\varphi)$ is the standard real $3\times 3$ matrix representation of SO(2) rotations around $\ff e_{z}$.
Hence, the first matrix element in Eq.\ (\ref{eq:lehmann}) can be written as
$\bra{\Psi_{0}} U_{\rm R} U_{\rm R}^\dagger s_{i\alpha} U_{\rm R} U_{\rm R}^\dagger \ket{\Psi_{n}} = 
\sum_{\beta} R_{\alpha\beta}(\varphi) e^{i \phi_{0}} \bra{\Psi_{0}} s_{i\beta} \ket{\Psi_{n}} 
e^{-i \phi_{n}}$. 
The phase factors cancel with those from the second matrix element, and we thus find:
$\chi_{ii',\alpha\alpha'}(\omega) = \sum_{\beta\beta'} R_{\alpha\beta}(\varphi) \chi_{ii', \beta\beta'}(\omega) R^{T}_{\beta'\alpha'}(\varphi)$, i.e., $[\underline{\chi}_{ii'}(\omega), \underline{R}(\varphi)]=0$ for all $i,i'$ and $\varphi$.
It is easily verified directly that this implies 
\be
\underline{\chi}_{ii'}(\omega)
= 
\begin{pmatrix} 
\chi_{ii',xx}(\omega) & \chi_{ii',xy}(\omega) & 0 \\ -\chi_{ii',xy}(\omega) & \chi_{ii',xx}(\omega) & 0 \\ 0 & 0 & \chi_{ii',zz}(\omega) \\
\end{pmatrix}
\: ,
\labeq{chiso2}
\ee
i.e., there are only 3 independent entries for each pair $i,i'$
(note that $R$ is reducible, and furthermore Schur's lemma does not apply to representations over $\mathbb{R}$).

\mbox{} \newline
\paragraph{\color{blue} 
Section B: Spatial symmetries.} 

With \refeq{chileh}, we immediately see that the spin-Berry curvature is related to the spin susceptibility via
\be
\Omega_{mm',\alpha\alpha'}
=
- i J^{2}
\frac{\partial}{\partial \omega} \chi_{i_{m} i_{m'},\alpha\alpha'}(\omega) \Big|_{\omega =0}
\labeq{omchi1}
\ee
up to correction terms of order $J^{3}$, see \refeq{omchi}. 
Therefore, the same reasoning as above can be applied to the spin-Berry curvature tensor and yields the same result for its structure:
\be
\underline{\Omega}_{mm'}
= 
\begin{pmatrix} 
\Omega_{mm',xx} & \Omega_{mm',xy} & 0 \\ -\Omega_{mm',xy} & \Omega_{mm',xx} & 0 \\ 0 & 0 & \Omega_{mm',zz} \\
\end{pmatrix}
\: .
\labeq{omso2}
\ee

In addition, for a given pair of sites $i_{m}$ and $i_{m'}$, we may consider a combined transformation $T \circ I$, composed of the space inversion $\ff R_{i} \mapsto \ff R_{i_{m}} - \ff R_{i}$ with respect to $i_{m}$ followed by the translation $\ff R_{i} \mapsto \ff R_{i} + (\ff R_{i_{m}} - \ff R_{i_{m'}})$ with the translation vector $\ff R_{i_{m}} - \ff R_{i_{m'}}$.
$T \circ I$ is a discrete symmetry of the hypercubic lattice and interchanges $i_{m}$ with $i_{m'}$. 
This implies that the Hamiltonian commutes with the standard unitary (and also Hermitian) representation $U_{\rm TI}$ of $T \circ I$. 

For $i_{m}$ and $i_{m'}$ in the same sublattice, the symmetry-broken ground state is an eigenstate of $U_{\rm TI}$ as well. 
Analogously to the SO(2) spin-rotation symmetry discussed above, we can thus immediately see from the analysis of the matrix elements in Eq.\ (\ref{eq:lehmann}) that $\Omega_{mm',\alpha\alpha'} = \Omega_{m'm,\alpha\alpha'}$.
For the spin-Berry curvature we have the additional antisymmetry, $\Omega_{mm',\alpha \alpha'}=-\Omega_{m' m,\alpha' \alpha}$, which follows from Eq.\ (\ref{eq:curv}). 
With this we get $\Omega_{mm',\alpha\alpha'} = - \Omega_{mm',\alpha'\alpha}$, i.e., for each pair $m,m'$, the spin-Berry curvature tensor is antisymmetric in the indices $\alpha, \alpha'$ separately.
Hence, with \refeq{omso2}, we see that only the elements $\Omega_{mm',xy} = - \Omega_{mm',yx}$ can be nonzero.
Analogously, for the spin susceptibility, the $T \circ I$ symmetry of the Hamiltonian implies $\chi_{ii',\alpha\alpha'}(\omega) = \chi_{i'i,\alpha\alpha'}(\omega)$. 
With the additional symmetry, $\chi_{ii',\alpha\alpha'}(0) = \chi_{i'i,\alpha'\alpha}(0)$, which follows from \refeq{chileh}, this implies that the susceptibility matrix \refeq{chiso2} is diagonal for $\omega=0$.

For $i_{m}$ and $i_{m'}$ in different sublattices, we concatenate the transformation $T \circ I$ with a flip $F$ of the $z$-component of all spins. 
This is unitarily represented by $U_{\rm F}$, which is defined via $U^{\dagger}_{\rm F} c_{i\uparrow} U_{\rm F} = c_{i\downarrow}$ and $U^{\dagger}_{\rm F} c_{i\downarrow} U_{\rm F} = c_{i\uparrow}$.
We have $[U_{\rm F}, U_{\rm TI} ] =0$ and $[U_{\rm F}, H_{\rm el}] =0$.
The symmetry-broken ground state and the corresponding excited states are eigenstates of $U \equiv U_{\rm TI} U_{\rm F}$. 
Hence, we find for the $\alpha=\alpha'$ matrix elements
\ba
&& \bra{\Psi_{0}} s_{i_{m} \alpha} \ket{\Psi_{n}} \bra{\Psi_{n}} s_{i_{m'} \alpha} \ket{\Psi_{0}} 
\nonumber \\
&=&
\bra{\Psi_{0}} U^{\dagger} s_{i_{m} \alpha} U \ket{\Psi_{n}} \bra{\Psi_{n}} U^{\dagger} s_{i_{m'} \alpha} U \ket{\Psi_{0}} 
\nonumber \\
&=&
\bra{\Psi_{0}} s_{i_{m'} \alpha} \ket{\Psi_{n}} \bra{\Psi_{n}} s_{i_{m} \alpha} \ket{\Psi_{0}} 
\:, 
\ea
and thus $\Omega_{mm',\alpha\alpha} = \Omega_{m'm,\alpha\alpha}$. 
With the antisymmetry of the full tensor, $\Omega_{mm', \alpha \alpha'}=-\Omega_{m'm, \alpha' \alpha}$, we thus find
$\Omega_{mm',\alpha\alpha} = 0$.
On the other hand, 
\ba
&& \bra{\Psi_{0}} s_{i_{m} x} \ket{\Psi_{n}} \bra{\Psi_{n}} s_{i_{m'} y} \ket{\Psi_{0}} 
\nonumber \\
&=&
\bra{\Psi_{0}} U^{\dagger} s_{i_{m} x} U \ket{\Psi_{n}} \bra{\Psi_{n}} U^{\dagger} s_{i_{m'} y} U \ket{\Psi_{0}} 
\nonumber \\
&=&
- \bra{\Psi_{0}} s_{i_{m'} x} \ket{\Psi_{n}} \bra{\Psi_{n}} s_{i_{m} y} \ket{\Psi_{0}} 
\: , 
\ea
since $U_{\rm F}^{\dagger} s_{ix} U_{\rm F} = s_{ix}$ but $U_{\rm F}^{\dagger} s_{iy} U_{\rm F} = - s_{iy}$. 
This implies $\Omega_{mm',xy} = -\Omega_{m'm,xy}$, and with the antisymmetry of the full tensor we find $\Omega_{mm',xy} = \Omega_{mm',yx}$. 
Together with \refeq{omso2}, we see that $\Omega_{mm',xy} = 0$, and hence the matrix $\underline{\Omega}_{mm'} = 0$ in \refeq{omso2}.

Summing up, for arbitrary sites $i_{m}$ and $i_{m'}$ we have 
\be
\underline{\Omega}_{mm'}
= 
\begin{pmatrix} 
0 & \Omega & 0 \\ - \Omega & 0 & 0 \\ 0 & 0 & 0 \\
\end{pmatrix}
\: ,
\labeq{omega}
\ee
and hence the spin-Berry curvature is fixed by a single real number $\Omega \equiv \Omega_{mm',xy}=\Omega_{m'm,xy}$. 
Furthermore, $\Omega = 0$ if $i_{m}, i_{m'}$ belong to different sublattices.

\mbox{} \newline
\paragraph{\color{blue} 
Section C: Random phase approximation.} 

The random phase approximation (RPA) represents a standard weak-coupling approach to the magnetic susceptibility, see, e.g., Refs.\ \onlinecite{RKEH12,DT21}. 
It can be motivated in various ways, for example, via a partial diagrammatic summation.
In general, the RPA Luttinger-Ward functional $\Phi[\ff G]$ \cite{LW60,RHT+18} is given as the sum of the two closed and self-consistently  renormalized first-order diagrams, i.e., by the Hartree and the Fock diagram. 
For the Hubbard model the Fock diagram vanishes such that we are left with
\be
\Phi[\ff G] = U \sum_{i} \frac{1}{\beta^2} \sum_{n, n'} G_{ii,\uparrow}(i\omega_n) G_{ii,\downarrow}(i\omega_{n'})
\: .
\ee
Here, $i$ runs over the sites of the hypercubic lattice, $\sigma=\uparrow, \downarrow$ refers to the spin projection relative to the $z$ axis, $n$ labels the fermionic Matsubara frequencies $i \omega_{n}$, and $\beta$ is the inverse temperature. 
Computations are done in the zero-temperature limit $1/\beta \to 0$, which is taken at the end.
Furthermore, $G_{ii,\sigma}$ denotes the local one-particle Green's function at site $i$ in the symmetry-broken AF state, as obtained within the self-consistent Hartree-Fock approximation.
The Hartree-Fock self-energy is generated by the Luttinger-Ward functional: 
$\Sigma_{ii,\sigma}(i \omega_n) = \beta \delta \Phi / \delta G_{ii\sigma}(i \omega_n) = U \beta^{-1} \sum_{n} G_{ii,-\sigma} (i\omega_{n}) = U \langle c^{\dagger}_{i-\sigma} c_{i-\sigma} \rangle$.

On the two-particle level, the RPA yields a local and frequency-independent irreducible vertex 
\be
\Gamma^{\rm (loc)}_{\sigma, -\sigma}(i\omega_n, i\omega_{n'}) = \beta^2 \frac{\delta^{2} \Phi}{\delta G_{ii,-\sigma}(i \omega_n) \delta G_{ii,\sigma}(i \omega_{n'})} = U
\: . 
\ee
This means that there is no feedback of two-particle correlations on the single-particle Green's function. 
The structureless vertex allows us to easily get the transversal magnetic susceptibility 
$\chi_{rs, +-}(\ff k, \omega) = \langle \langle s_{r,\ff k}^{+} ; s_{s,\ff k}^{-} \rangle \rangle_{\omega}$ 
as the solution of a strongly simplified Bethe-Salpether equation in the particle-hole channel:
\be
\underline{\chi}_{+-}(\ff k, i\nu_{n}) = \underline{\chi}^{(0)}_{+-}(\ff k, i\nu_{n}) + \underline{\chi}^{(0)}_{+-} (\ff k, i \nu_{n}) U \underline{\chi}_{+-}(\ff k, i \nu_{n}) 
\: .
\labeq{bs}
\ee  
Here, $\underline{\chi}_{+-}$ is a $2\times 2$ matrix in the sublattice degrees of freedom, and $\underline{\chi}^{(0)}_{+-}$ the bare susceptibility matrix, which is computed with the Hartree-Fock one-particle propagators.
The equation is diagonal in the wave vectors $\ff k$ of the first magnetic Brillouin zone and in the bosonic Matsubara frequencies $i\nu_{n}$.
The transversal susceptibility $\chi_{+-}$ is related to the susceptibility tensor $\chi_{\alpha\alpha'}$ introduced in Eqs.\ (\ref{eq:chi1}) and (\ref{eq:chileh}), via $\chi_{+-} = 2 (\chi_{xx} - i \chi_{xy})$.
From the renormalized zeroth-order diagram, we get the Hartree-Fock susceptibility in \refeq{bs} as
\ba
{\chi}^{(0)}_{rs,+-} (\ff k, i\nu_n)
&=& 
\frac{-1}{L} \frac{1}{\beta} \sum_{\ff q, n'} G_{sr,\uparrow}(\ff q, i\omega_{n'}) 
\nonumber \\
&\times & G_{rs, \downarrow}(\ff q+\ff k, i\nu_{n} + i\omega_{n'}) 
\: .
\labeq{chi0}
\ea
After performing the summation over the fermionic frequencies $i\omega_{n}$ analytically, we can replace $i\nu_{n} \mapsto \nu + i \eta$ to find the retarded susceptibility on the real-frequency axis. 
The frequency derivative in \refeq{omchi1} is done numerically. 
In the thermodynamical limit, the $\ff q$ sum over the magnetic Brillouin zone in \refeq{chi0} can be converted into a $\ff q$-space integration. 
The latter is computed in two or three dimensions via a standard adaptive $\ff q$-space integration technique for arbitrary $\nu \in \mathbb{R}$ and for each allowed wave vector $\ff k$ in the magnetic Brillouin zone of a finite lattice with $L$ sites and periodic boundary conditions. 
Practical computations are performed at a finite Lorentzian broadening parameter $\eta>0$ replacing the infinitesimal $\eta$, and convergence with respect to $L$ is controlled by runs for different system sizes $L$. 
The main numerical error is due to extrapolation of the data for $\eta \to 0$.

\mbox{} \newline
\paragraph{\color{blue} 
Section D: Magnon spectrum of an antiferromagnet.} 

In the strong-$U$ limit of the Hubbard model, the low-energy physics is captured by the $s=1/2$ antiferromagnetic Heisenberg model
\be
H = J_{\rm H} \sum_{\langle ij \rangle} \left(\frac{1}{2}(s^{+}_is^{-}_j + s^{-}_is^{+}_j) + \Delta s^{z}_i s^{z}_j\right)
\labeq{heisham}
\ee
with $J_{\rm H}=4t^{2}/U$ and $\Delta=1$. 
An anisotropy parameter $\Delta > 1$ can be used to discuss the effect of opening a gap in the dispersion.
The sum runs over all nearest-neighbor pairs $\langle ij \rangle$.

We apply the standard Holstein-Primakoff transfomation for the model on the bipartite hypercubic lattice with dimension $D$. 
For sites $i$ in sublattice A, the spin operators are expressed in terms of bosonic annihilators and creators, i.e., 
\ba
s^{z}_i &=& s - a^\dagger_ia_i \: ,
\nonumber \\
s^{+}_i &=& \sqrt{2s} \sqrt{1 - \frac{\hat{n}_i}{2s}} a_i  \: , 
\nonumber \\
s^{-}_i &=& \sqrt{2s} a^\dagger_i \sqrt{1 - \frac{\hat{n}_i}{2s}}  \: , 
\ea
while for sites $j \in B$
\ba
s^{z}_j &=& -s + b^\dagger_jb_j \: , 
\nonumber \\
s^{+}_j &=& \sqrt{2s} b^\dagger_j \sqrt{1 - \frac{\hat{n}_j}{2s}}  \: , 
\nonumber \\
s^{-}_j &=& \sqrt{2s} \sqrt{1 - \frac{\hat{n}_j}{2s}} b_j  \: .
\ea
The transformed Hamiltonian reads
\be
H 
= J_{\rm H} s \sum_{\langle ij \rangle} [  (a_ib_j + a^\dagger_ib^\dagger_j) + \Delta (a^\dagger_ia_i + b^\dagger_jb_j) ] 
- \frac{L}{2} z J_{\rm H} \Delta s^2\; , 
\ee
where $z=2D$ is the coordination number, and $L$ is the number of lattice sites.
Quartic and higher-order magnon interaction terms resulting for the expansion of the square root have been disregarded.

We drop the additive energy constant and block-diagonalize $H$ via Fourier transformation:
\ba
a_i = \frac{1}{\sqrt{L/2}} \sum_{\boldsymbol{k}} e^{-i\boldsymbol{k}\boldsymbol{R}_i} a_{\boldsymbol{k}}
\: , \quad b_j = \frac{1}{\sqrt{L/2}} \sum_{\boldsymbol{k}} e^{i\boldsymbol{k}\boldsymbol{R}_j} b_{\boldsymbol{k}} 
\: .
\nonumber \\ 
\ea
Here, $\ff R_{i}$ are the translation vectors of the magnetic A sublattice (the same for $\ff R_{j}$ and the B sublattice), consisting of $L/2$ unit cells, and $\ff k$ is an allowed wave vector of the first magnetic Brillouin zone (mBz).
Defining $\gamma_{\ff k} \equiv \sum_{\ff \delta} \cos(\ff k \ff \delta)$ with nearest-neighbor vectors $\ff \delta$, the Fourier-transformed model reads as
\be
H 
= 
J_{\rm H}s \sum_{\boldsymbol{k}} [ \gamma_{\boldsymbol{k}} (a_{\boldsymbol{k}}b_{\boldsymbol{k}} + b^\dagger_{\boldsymbol{k}}a^\dagger_{\boldsymbol{k}}) + z\Delta (a_{\boldsymbol{k}}a^\dagger_{\boldsymbol{k}} + b^\dagger_{\boldsymbol{k}}b_{\boldsymbol{k}}) ] 
\ee
and can be diagonalized by Bogoliubov transformation
\be
a_{\boldsymbol{k}}  = u_{\boldsymbol{k}} \alpha_{\boldsymbol{k}} + v_{\boldsymbol{k}} \beta^\dagger_{\boldsymbol{k}} 
\: , \quad
b_{\boldsymbol{k}}  = u_{\boldsymbol{k}} \beta_{\boldsymbol{k}} + v_{\boldsymbol{k}} \alpha^\dagger_{\boldsymbol{k}}
\ee
with real coefficients $u_{\ff k}$ and $v_{\ff k}$.
We require
\be
u^2_{\boldsymbol{k}} - v^2_{\boldsymbol{k}} = 1 \: , 
\labeq{cond1}
\ee
to ensure that $\alpha_{\boldsymbol{k}}$ and $\beta_{\boldsymbol{k}}$ satisfy bosonic commutation relations, as well as
\be
2 z \Delta u_{\boldsymbol{k}}v_{\boldsymbol{k}} + \gamma_{\boldsymbol{k}}(u^2_{\boldsymbol{k}} + v^2_{\boldsymbol{k}}) 
\stackrel{!}{=} 
0 \: , 
\labeq{cond2}
\ee
as usual, to get the Hamiltonian to the form
\be
H = J_{\rm H}s \sum_{\boldsymbol{k}} \omega(\boldsymbol{k}) 
(\alpha^\dagger_{\boldsymbol{k}} \alpha_{\boldsymbol{k}} + \beta^\dagger_{\boldsymbol{k}} \beta_{\boldsymbol{k}})
\: , 
\ee
where we again dropped an unimportant constant energy term.
The magnon spectrum consists of two degenerate branches with dispersion $J_{\rm H}s \, \omega(\ff k)$ given by
\be
\omega(\boldsymbol{k}) 
=
\sqrt{(z \Delta)^2 - \gamma^2_{\ff k}} 
=
\sqrt{z^{2} \Delta^{2} - \left(\sum_{\ff \delta} \cos(\ff k \ff \delta)\right)^{2}}
\: .
\labeq{omk}
\ee
Close to $\ff k=0$ and in the isotropic case $\Delta=1$, the dispersion $J_{\rm H}s\, \omega(\ff k)$ is linear, 
\be
\omega(\ff k) = 2 \sqrt{D} k + \ca O(k^{3})
\: ,
\labeq{linearom}
\ee 
while for $\Delta > 1$ the spectrum is gapped, and $\omega(\ff k) = 2S \sqrt{\Delta^{2}-1} + \ca O(k^{2})$.

From the conditions \refeq{cond1} and \refeq{cond2}, we can deduce the well-known results
\ba
u^2_{\boldsymbol{k}} 
&=& 
\frac{1}{2} \left(\frac{z \Delta}{\sqrt{(z \Delta)^2 - \gamma^2_{\boldsymbol{k}}}} + 1\right) 
= 
\frac{1}{2} \left(\frac{z \Delta}{\omega(\boldsymbol{k})} + 1\right) 
\: , \nonumber \\
v^2_{\boldsymbol{k}} 
&=& \frac{1}{2} \left(\frac{z \Delta}{\sqrt{(z \Delta)^2 - \gamma^2_{\boldsymbol{k}}}} - 1\right) 
= 
\frac{1}{2} \left(\frac{z \Delta}{\omega(\boldsymbol{k})} - 1\right)
\ea
and 
\be
u_{\boldsymbol{k}} v_{\boldsymbol{k}} = -\frac{\gamma_{\boldsymbol{k}}}{2\sqrt{(z \Delta)^2 - \gamma^2_{\boldsymbol{k}}}} 
= 
-\frac{\gamma_{\boldsymbol{k}}}{2\omega(\boldsymbol{k})}
\: ,
\ee
see Refs.\ \onlinecite{And52,HP40}.

\mbox{} \newline
\paragraph{\color{blue} 
Section E: Computing the spin-Berry curvature from the magnon Hamiltonian.} 

The contribution of the magnon excitations to the spin-Berry curvature is obtained from
\be
\Omega_{mm', \alpha \alpha'} = -2J^2 \mbox{Im}\sum_{\boldsymbol{k}, \eta=1,2} \frac{\bra{0} s^{\alpha}_{i_{m}} \ket{\boldsymbol{k},\eta}\bra{\boldsymbol{k},\eta} s^{\alpha'}_{i_{m'}} \ket{0}}{(E_0 - E_{\boldsymbol{k}})^2}
\:, 
\ee
where $\ket{\boldsymbol{k},1} \equiv \alpha^\dagger_{\boldsymbol{k}} \ket{0}$ and $\ket{\boldsymbol{k},2} \equiv \beta^\dagger_{\boldsymbol{k}} \ket{0}$ are the single-magnon states. 
Following \refeq{omega}, it is sufficient to compute $\Omega_{mm'} = \Omega_{mm',xy} = - \Omega_{mm',yx}$. 
Furthermore, $\Omega_{mm'}\ne 0$ only for $i_{m}$ and $i_{m'}$ in the same sublattice, as also argued in section B.
Expressing the spin components in terms of the Bogoliubov operators, 
\ba
s^{+}_i &=& \sqrt{2s} \frac{1}{\sqrt{L/2}}\sum_{\boldsymbol{k}} e^{-i\boldsymbol{k}\boldsymbol{R}_i} (u_{\boldsymbol{k}} \alpha_{\boldsymbol{k}} + v_{\boldsymbol{k}} \beta^\dagger_{\boldsymbol{k}})
\nonumber \\
s^{-}_i &=& \sqrt{2s} \frac{1}{\sqrt{L/2}} \sum_{\boldsymbol{k}} e^{i\boldsymbol{k}\boldsymbol{R}_i} (u_{\boldsymbol{k}} \alpha^\dagger_{\boldsymbol{k}} + v_{\boldsymbol{k}} \beta_{\boldsymbol{k}}) 
\nonumber \\
s^{+}_j &=& \sqrt{2s} \frac{1}{\sqrt{L/2}}\sum_{\boldsymbol{k}} e^{-i\boldsymbol{k}\boldsymbol{R}_j} (u_{\boldsymbol{k}} \beta^\dagger_{\boldsymbol{k}} + v_{\boldsymbol{k}} \alpha_{\boldsymbol{k}})
\nonumber \\
s^{-}_j &=& \sqrt{2s} \frac{1}{\sqrt{L/2}} \sum_{\boldsymbol{k}} e^{i\boldsymbol{k}\boldsymbol{R}_j} (u_{\boldsymbol{k}} \beta_{\boldsymbol{k}} + v_{\boldsymbol{k}} \alpha^\dagger_{\boldsymbol{k}})
\: ,
\ea
we find 
\ba
\bra{0} s^{x}_{i} \ket{\boldsymbol{k},1} \bra{\boldsymbol{k},1} s^{y}_{i^\prime} \ket{0} 
&=& 
i \frac{s}{L} u^2_{\boldsymbol{k}} e^{-i\boldsymbol{k}(\boldsymbol{R}_i - \boldsymbol{R}_{i^\prime})} 
\: ,
\nonumber \\
\bra{0} s^{x}_{i} \ket{\boldsymbol{k},2} \bra{\boldsymbol{k},2} s^{y}_{i^\prime} \ket{0} 
&=& 
-i \frac{s}{L} v^2_{\boldsymbol{k}} e^{i\boldsymbol{k}(\boldsymbol{R}_i - \boldsymbol{R}_{i^\prime})} 
\ea
for $i$, $i'$ in sublattice A, and
\ba
\bra{0} s^{x}_{j} \ket{\boldsymbol{k},1} \bra{\boldsymbol{k},1} s^{y}_{j^\prime} \ket{0} 
&=& 
i \frac{s}{L} v^2_{\boldsymbol{k}} e^{-i\boldsymbol{k}(\boldsymbol{R}_j - \boldsymbol{R}_{j^\prime})} 
\: ,
\nonumber \\
\bra{0} s^{x}_{j} \ket{\boldsymbol{k},2} \bra{\boldsymbol{k},2} s^{y}_{j^\prime} \ket{0} 
&=& 
-i \frac{s}{L} u^2_{\boldsymbol{k}} e^{i\boldsymbol{k}(\boldsymbol{R}_j - \boldsymbol{R}_{j^\prime})} 
\ea
for $j$, $j'$ in sublattice B. 
This implies
\ba
\mbox{Im}{\sum_{\eta=1,2} \bra{0} s^{x}_{i} \ket{\boldsymbol{k}, \eta} \bra{\boldsymbol{k}, \eta} s^{y}_{i^\prime} \ket{0}} 
&=&  
\frac{s}{L} \cos(\boldsymbol{k}(\boldsymbol{R}_i - \boldsymbol{R}_{i^\prime})) 
\nonumber \\
\mbox{Im}{\sum_{\eta =1,2} \bra{0} s^{x}_{j} \ket{\boldsymbol{k}, \eta} \bra{\boldsymbol{k}, \eta} s^{y}_{j^\prime} \ket{0}} 
&=& 
-\frac{s}{L} \cos(\boldsymbol{k}(\boldsymbol{R}_j - \boldsymbol{R}_{j^\prime}))
\: , 
\nonumber \\
\ea
and finally we get
\be
\Omega_{mm^\prime} 
= \mp
\frac{1}{s} \frac{J^2}{J_{\rm H}^2} \frac{2}{L} \sum^{\rm mBz}_{\boldsymbol{k}} \frac{\cos(\boldsymbol{k}(\boldsymbol{R}_{i_{m}} -\boldsymbol{R}_{i_{m'}}))}{\omega(\boldsymbol{k})^2}
\: ,
\labeq{omresult}
\ee
where the upper sign refers to $i_{m},i_{m'}$ in sublattice A and the lower for $i_{m},i_{m'}$ in sublattice B.
Recall that $\Omega_{mm^\prime}=0$ if $i_{m},i_{m'}$ belong to different sublattices.

Eq.\ (\ref{eq:omresult}) can be evaluated numerically. 
For $D=3$, for example, we find
\be
\Omega_{\rm loc} \approx -0.084 \frac{J^2}{J_{\rm H}^{2}} \: 
\ee
for the local element of the SBC with $i_{m}=i_{m'}$ in sublattice A.

\mbox{} \newline
\paragraph{\color{blue} 
Section F: Different dimensions and distance dependence.} 

In the thermodynamic limit $L\to \infty$ (and in the isotropic case $\Delta=1$), the convergence of the resulting integral in \refeq{omresult} over the magnetic Brillouin zone decisively depends on the lattice dimension $D$. 
We consider the critical contribution of the long-wave-length magnons by integrating over a $D$-dimensional ball around $\ff k=0$ with small cutoff radius $k_{\rm c}$, such that we can make use of \refeq{linearom}, i.e., of the linearity and isotropy of the magnon dispersion for $k\to 0$:
\be
\Omega_{mm'}
\sim
\lim_{\kappa \to 0}
\int_{\kappa}^{k_{\rm c}} dk \, k^{D-1}
\frac{1}{\omega(\boldsymbol{k})^2}
\propto 
\lim_{\kappa \to 0}
\int_{\kappa}^{k_{\rm c}} dk \, k^{D-1}
\frac{1}{k^{2}}
\: .
\ee
This yields
\be
\Omega_{mm'}
\sim 
\begin{cases}
\kappa^{D-2} \quad \text{ for } D \geq 3 \\
\ln \kappa \quad \text{ for } D = 2 \\
1/\kappa \quad \text{ for } D = 1
\end{cases}
\: . 
\ee
For $\kappa \to 0$, the spin-Berry curvature diverges for $D=1$ and $D=2$. 
We conclude that a meaningful theory is obtained in dimensions $D \ge 3$ only. 

The magnitude of the spin-Berry curvature decreases with increasing distance $\ff R \equiv \ff R_{i_{m}} - \ff R_{i_{m'}}$. 
For $D\ge 3$ its dependence in the large-$R$ limit is governed by long-wave-length magnon excitations, and we have:
\be
\Omega(R)
\propto
\int_{0}^{k_{c}} dk \, k^{D-1}
\int d\Omega
\frac{\cos(k R \cos \theta)}{k^{2}}
\: , 
\ee
where $\int d\Omega$ denotes the surface integral over the $(D-1)$-dimensional unit sphere, and $\theta$ the angle between $\ff k$ and $\ff R$. 
Furthermore, we made use of \refeq{linearom} for $k$ smaller than the cutoff $k_{\rm c}$.
We note that the distance dependence at large $R$ is isotropic.
Substituting $kR \to k$ in the one-dimensional $k$ integral immediately yields
\be
\Omega(R) \propto \frac{1}{R^{D-2}}
\: .
\ee
For $D=3$, we have $\Omega(R) \propto 1/R$. 
In the infinite-$D$ limit, we expect a local spin-Berry curvature 

To compute the local element $m=m'$ of the spin-Berry curvature \refeq{omresult} in this limit, we start from the representation 
\be
\Omega_{\rm loc} 
= 
-\frac{1}{s} \frac{J^2}{J_{\rm H}^2} 
\int_{-\infty}^{\infty} dx \,  \rho_{D}(x)
\frac{1}{z^{2} \Delta^{2} - D x^{2}}
\: , 
\ee
where, for dimension $D$, we have defined the density function
\be
\rho_{D}(x) = \frac{2}{L} \sum_{\ff k}^{\rm mBz} \delta(x-\gamma_{\ff k} / \sqrt{D}) \: ,
\ee
and where we have used \refeq{omk}.
We have $z=2D$ for the $D$-dimensional hypercubic lattice and, in the Heisenberg limit of the Hubbard model, $J_{\rm H}=4t^{2}/U = 4t^{\ast}{}^{2}/DU$, when using the scaling $t=t^{\ast}/\sqrt{D}$ with $t^{\ast}=\mbox{const}$. 
In the limit $D\to \infty$, this scaling of the hopping ensures that the kinetic energy of the Hubbard model remains nontrivial and balances the interaction term \cite{MH89b}. 
Moreover, the density function converges to a Gaussian \cite{MH89b}:
\be
\rho_{D}(x) \to \rho_{\infty}(x) = \frac{1}{2\sqrt{\pi}} \exp\left( - \frac{x^{2}}{4} \right) \: . 
\ee
In the Heisenberg limit and with the scaled hopping, we thus have
\be
\Omega_{\rm loc} (D)
= 
-\frac{1}{s} \frac{J^2 U^{2}}{16 t^{\ast}{}^{4}}
\int_{-\infty}^{\infty} 
\frac{dx \rho_{D}(x)}{4 \Delta^{2} - x^{2} / D}
\: , 
\ee
which for $D\to \infty$, and assuming $s=1/2$ and $\Delta = 1$ converges to 
\be
\Omega_{\rm loc} (\infty)
= 
- \frac{1}{32t^{\ast}{}^{4}} J^{2} U^{2}
\: .
\ee
This represents the mean-field value of the (lcoal) spin-Berry curvature in the antiferromagnetic state at large $U$.

To compare with the result obtained for $D=3$, we must use the same scaling of the hopping. 
This yields
\be
\Omega_{\rm loc} (3) \approx -0.084 \frac{J^2 U^{2}}{16 {t^{\ast}}^{4}} D^{2} \Bigg|_{D=3} \approx 1.51 \cdot \Omega(\infty) . 
\ee
For lattice dimensions $D > 3$ we find:
$\Omega_{\rm loc} (4) \approx 1.22 \, \Omega(\infty)$, 
$\Omega_{\rm loc} (5) \approx 1.16 \, \Omega(\infty)$,
$\Omega_{\rm loc} (6) \approx 1.12 \, \Omega(\infty)$.
Hence, given the standard scaling of the hopping with $D$, the absolute value of $\Omega_{\rm loc}(D)$ increases with decreasing $D$ and finally, for $D=2$ diverges.

Finally, when addressing the dimensional crossover \cite{BG90,MSS92,MSS93}, we consider the Heisenberg model given by \refeq{heisham} again, but with spatially anisotropic nearest-neighbor exchange couplings $J_{\rm H} \equiv J_{\rm H,x} = J_{\rm H,y} \ge J_{\rm H,z}$. 
Proceeding analogously to Sec.\ C, one ends up with a modified magnon dispersion only:
\be
  \omega(\boldsymbol{k}) = \sqrt{
  (z_{\text{eff}} \Delta)^2 - \gamma^{\prime 2}_{\boldsymbol{k}} 
  } 
  \: . 
\ee
Here, we have defined an effective coordination number $z_{\text{eff}} = 2 (J_{\rm H, x} + J_{\rm H, y} + J_{\rm H,z}) / J_{\rm H,x}$. 
Furthermore, $\gamma^{\prime}_{\boldsymbol{k}} := 2(J_{\rm H,x} \cos k_x + J_{\rm H,y} \cos k_y + J_{\rm H,z} \cos k_z)/J_{\rm H,x}$.

\begin{widetext}

\mbox{} \newline
\paragraph{\color{blue} 
Section G: Spin dynamics.} 

The equations of motion \refeq{eom} for the classical spins comprise the conventional (Hamiltonian) and the geometrical spin torque, see \refeq{geo}. 
In the weak-$J$ limit, the former results from the local direct exchange $J$ as well as from the indirect RKKY-type exchange. 
We have:
\be
\dot{\boldsymbol{S}}_m 
= 
J \langle \boldsymbol{s}_{i_m} \rangle^{(0)} \times \boldsymbol{S}_m 
+ 
J^2 \sum_{m^\prime} \underline{\chi}_{i_{m} i_{m'}}(0) \boldsymbol{S}_{m^\prime} \times \boldsymbol{S}_m 
+ 
\sum_{\alpha} \sum_{m'\alpha'} \Omega_{m'm,\alpha'\alpha}(\ff S) \dot{S}_{m'\alpha'} \ff e_{\alpha} \times \boldsymbol{S}_m
\; , 
\labeq{eomm}
\ee
where $\langle \dots \rangle^{(0)}$ denotes the expectation value at $J=0$. 
For the non-vanishing components of the spin susceptibility and of the spin-Berry curvature on sublattice A we have
\be
\chi_{ii'} \equiv \chi_{ii',xx}(0) = \chi_{ii',yy}(0) 
= 
-\frac{z \Delta}{J_{\rm H}} \frac{2}{L} \sum_{\boldsymbol{k}}^{\rm mBz} \frac{\cos{\boldsymbol{k}(\boldsymbol{R}_i - \boldsymbol{R}_{i^\prime})}}{\omega(\boldsymbol{k})^2}
\ee
and 
\be
\Omega_{mm'} \equiv \Omega_{mm',xy}=-\Omega_{mm',yx} 
= 
- \frac{1}{s}\frac{J^2}{J_{\rm H}^2} \frac{2}{L}
\sum_{\boldsymbol{k}}^{\rm mBz} \frac{\cos(\boldsymbol{k}(\boldsymbol{R}_{i_{m}} -\boldsymbol{R}_{i_{m'}})}{\omega(\boldsymbol{k})^2}
\: .
\ee

Specializing \refeq{eomm} for $M=1$, i.e., for a single classical spin, we get
\be
\dot{\boldsymbol{S}}_1 = \ff T^{\rm (H)}_{1} \times \boldsymbol{S}_1
+ \Omega_{11} (\ff e_z \times \dot{\boldsymbol{S}}_1) \times \boldsymbol{S}_1 
\: ,	
\ee
where 
\be
  \ff T^{\rm (H)}_{1} = J \langle \boldsymbol{s}_{i_1} \rangle^{(0)} + J^2 \chi_{i_{1}i_{1}} (\ff e_z \times \boldsymbol{S}_m) \times \ff e_z
\: .
\ee
With
\be
\underline{T}^{\rm (H)}_{1} = \begin{pmatrix}
0 & - T^{\rm (H)}_{1,z} & T^{\rm (H)}_{1,y} \\
T^{\rm (H)}_{1,z} & 0 & - T^{\rm (H)}_{1,x} \\
-T^{\rm (H)}_{1,y} &T^{\rm (H)}_{1,x} & 0
\end{pmatrix}
\ee
the cross product can be written as a matrix-vector product,
$\ff T^{\rm (H)}_{1} \times \boldsymbol{S}_1 = \underline{T}^{\rm (H)}_{1} \boldsymbol{S}_1$, and the equation of motion reads:
\be		
  \dot{\boldsymbol{S}}_1 = \frac{1}{1 - \Omega_{11}S_{1z}} \underline{T}^{\rm (H)}_{1} \boldsymbol{S}_1 \: .
\ee
The classical spin undergoes a purely precessional dynamics around the $z$ axis, but with a renormalized precession frequency. 
The renormalization is due to the local spin-Berry curvature $\Omega_{\rm loc} = \Omega_{11}$ and is the strongest for $\Omega_{\rm loc} = \ca O(1)$. 
Right at $\Omega_{\rm loc} = 1/S_{1z}$, the precession frequency diverges. 
This implies that the spin dynamics is no longer adiabatic and the theory breaks down. 

In case of two classical spins, $M=2$, the equations of motion (\ref{eq:eomm}) can be cast into the form
\ba
\dot{\boldsymbol{S}}_1 
&=&
\boldsymbol{T}^{\text{(H)}}_1 \times \ff S_{1}
+
\boldsymbol{T}^{\text{(geo)}}_1 \times \ff S_{1}
\: , 
\nonumber \\
\dot{\boldsymbol{S}}_2 
&=&
\boldsymbol{T}^{\text{(H)}}_2 \times \ff S_{2}
+
\boldsymbol{T}^{\text{(geo)}}_2 \times \ff S_{2}
\: ,
\ea
where
\ba
\boldsymbol{T}^{\text{(H)}}_1 
&=& 
J \langle \boldsymbol{s}_{i_1} \rangle^{(0)} 
+
J^2 \chi_{i_{1}i_{1}} (\ff e_{z} \times \boldsymbol{S}_1) \times \ff e_{z} + J^2 \chi_{i_{1}i_{2}} (\ff e_{z} \times \boldsymbol{S}_2) \times \ff e_{z}
\: , \nonumber \\
\boldsymbol{T}^{\text{(H)}}_2 
&=& 
J \langle \boldsymbol{s}_{i_2} \rangle^{(0)} 
+
J^2 \chi_{i_{2}i_{2}}  (\ff e_{z} \times \boldsymbol{S}_2) \times \ff e_{z} + J^2 \chi_{i_{2}i_{1}} (\ff e_{z} \times \boldsymbol{S}_1) \times \ff e_{z}
\: .
\ea
and
\ba
\boldsymbol{T}^{\rm (geo)}_1
&=& 
\Omega_{11} (\ff e_z \times \dot{\boldsymbol{S}}_1) \times \boldsymbol{S}_1 
+ 
\Omega_{12} (\ff e_z \times \dot{\boldsymbol{S}}_2) \times \boldsymbol{S}_1 
\nonumber \\
\boldsymbol{T}^{\rm (geo)}_2
&=& 
\Omega_{22} (\ff e_z \times \dot{\boldsymbol{S}}_2) \times \boldsymbol{S}_2 
+ 
\Omega_{12} (\ff e_z \times \dot{\boldsymbol{S}}_1) \times \boldsymbol{S}_2
\: .
\ea
Here, we have assumed that the two spins couple to sites in the same sublattice, as otherwise the spin-Berry curvature vanishes.
The local spin-Berry curvature term can be treated in the same way as in the $M=1$ case, while the nonlocal term can be written as a matrix-vector product:
\ba
(1 - \Omega_{11} S_{1z})\dot{\boldsymbol{S}}_1 
&=& 
\ff T^{\rm (H)}_{1} \times \boldsymbol{S}_1 - \Omega_{12} \underline{\mathcal{A}}^{(z)}_1 \dot{\boldsymbol{S}}_2 
\; , \nonumber \\
(1 - \Omega_{22} S_{2z})\dot{\boldsymbol{S}}_2 
&=& 
\ff T^{\rm (H)}_{2} \times \boldsymbol{S}_2 - \Omega_{12} \underline{\mathcal{A}}^{(z)}_2 \dot{\boldsymbol{S}}_1
\; , 
\ea
with
\be
\underline{\mathcal{A}}^{(z)}_m 
= 
\begin{pmatrix}
-S_{mz} & 0 & 0\\
0 & -S_{mz} & 0\\
S_{mx} & S_{my} & 0
\end{pmatrix}
\: .
\ee
This allows us to cast the equations of motion into an explicit system of ordinary differential equations:
\be
\begin{pmatrix}
\dot{\boldsymbol{S}}_1\\
\dot{\boldsymbol{S}}_2
\end{pmatrix} 
=
\ca M^{-1}	
\begin{pmatrix}
\ff T_{1}^{\rm (H)} \times \boldsymbol{S}_1 \\
\ff T_{2}^{\rm (H)} \times \boldsymbol{S}_2
\end{pmatrix} 
\: .
\label{eq:eom2}
\ee
Here, the $6\times 6$ matrix
\be
\ca M
=
\begin{pmatrix}
(1 - \Omega_{11} S_{1z}) {\ff 1} & \Omega_{12} \underline{\mathcal{A}}^{(z)}_1 \\
\Omega_{12} \underline{\mathcal{A}}^{(z)}_2 & (1 - \Omega_{22}S_{2z}) {\ff 1}
\end{pmatrix}
\ee
is given in terms of the components of the spin-Berry curvature tensor.
\refeq{eom2} demonstrates that the effect of the geometrical spin torque is not simply additive and hence does not directly compete with the conventional spin torque, but enters the spin dynamics as a multiplicative (matrix) factor. 

The determinant of $\mathcal{M}$ can be computed analytically: 
\be
\det \mathcal{M}
= 
(1 - \Omega_{11} S_{1z}) (1 - \Omega_{11} S_{2z}) 
\big[
(1 - \Omega_{11} S_{1z}) (1 - \Omega_{22} S_{2z}) - \Omega_{12}^{2} S_{1z} S_{2z}
\big]^{2}
\: .
\ee
The theory breaks down if $\det \mathcal{M}=0$.
We consider $\det \ca M$ as a function of the local elements $\Omega_{\rm loc}=\Omega_{11}=\Omega_{22}$ and assume that the nonlocal elements are small, $\Omega_{\rm nonloc}=|\Omega_{12}| \ll \Omega_{\rm loc}$.
We immediately see that the zeros of $\det \ca M$ are of the order of unity. 
This implies that anomalous spin dynamics, which is substantially affected by the geometrical spin torque, is expected if $\Omega_{\rm loc} = \ca O(1)$ and thus close to, but yet different from the zeros of $\ca M$.

\end{widetext}


\begin{thebibliography}{63}
\expandafter\ifx\csname natexlab\endcsname\relax\def\natexlab#1{#1}\fi
\expandafter\ifx\csname bibnamefont\endcsname\relax
  \def\bibnamefont#1{#1}\fi
\expandafter\ifx\csname bibfnamefont\endcsname\relax
  \def\bibfnamefont#1{#1}\fi
\expandafter\ifx\csname citenamefont\endcsname\relax
  \def\citenamefont#1{#1}\fi
\expandafter\ifx\csname url\endcsname\relax
  \def\url#1{\texttt{#1}}\fi
\expandafter\ifx\csname urlprefix\endcsname\relax\def\urlprefix{URL }\fi
\providecommand{\bibinfo}[2]{#2}
\providecommand{\eprint}[2][]{\url{#2}}

\bibitem[{\citenamefont{Bertotti et~al.}(2009)\citenamefont{Bertotti,
  Mayergoyz, and Serpico}}]{BMS09}
\bibinfo{author}{\bibfnamefont{G.}~\bibnamefont{Bertotti}},
  \bibinfo{author}{\bibfnamefont{I.~D.} \bibnamefont{Mayergoyz}},
  \bibnamefont{and} \bibinfo{author}{\bibfnamefont{C.}~\bibnamefont{Serpico}},
  \emph{\bibinfo{title}{Nonlinear Magnetization Dynamics in Nanosystemes}}
  (\bibinfo{publisher}{Elsevier}, \bibinfo{address}{Amsterdam},
  \bibinfo{year}{2009}).

\bibitem[{\citenamefont{Koshibae et~al.}(2009)\citenamefont{Koshibae, Furukawa,
  and Nagaosa}}]{KFN09}
\bibinfo{author}{\bibfnamefont{W.}~\bibnamefont{Koshibae}},
  \bibinfo{author}{\bibfnamefont{N.}~\bibnamefont{Furukawa}}, \bibnamefont{and}
  \bibinfo{author}{\bibfnamefont{N.}~\bibnamefont{Nagaosa}},
  \bibinfo{journal}{Phys. Rev. Lett.} \textbf{\bibinfo{volume}{103}},
  \bibinfo{pages}{266402} (\bibinfo{year}{2009}).

\bibitem[{\citenamefont{Bhattacharjee et~al.}(2012)\citenamefont{Bhattacharjee,
  Nordstr\"om, and Fransson}}]{BNF12}
\bibinfo{author}{\bibfnamefont{S.}~\bibnamefont{Bhattacharjee}},
  \bibinfo{author}{\bibfnamefont{L.}~\bibnamefont{Nordstr\"om}},
  \bibnamefont{and} \bibinfo{author}{\bibfnamefont{J.}~\bibnamefont{Fransson}},
  \bibinfo{journal}{Phys. Rev. Lett.} \textbf{\bibinfo{volume}{108}},
  \bibinfo{pages}{057204} (\bibinfo{year}{2012}).

\bibitem[{\citenamefont{Evans et~al.}(2014)\citenamefont{Evans, Fan,
  Chureemart, Ostler, Ellis, and Chantrell}}]{EFC+14}
\bibinfo{author}{\bibfnamefont{R.~F.~L.} \bibnamefont{Evans}},
  \bibinfo{author}{\bibfnamefont{W.~J.} \bibnamefont{Fan}},
  \bibinfo{author}{\bibfnamefont{P.}~\bibnamefont{Chureemart}},
  \bibinfo{author}{\bibfnamefont{T.~A.} \bibnamefont{Ostler}},
  \bibinfo{author}{\bibfnamefont{M.~O.~A.} \bibnamefont{Ellis}},
  \bibnamefont{and} \bibinfo{author}{\bibfnamefont{R.~W.}
  \bibnamefont{Chantrell}}, \bibinfo{journal}{J. Phys.: Condens. Matter}
  \textbf{\bibinfo{volume}{26}}, \bibinfo{pages}{103202}
  (\bibinfo{year}{2014}).

\bibitem[{\citenamefont{Sayad and Potthoff}(2015)}]{SP15}
\bibinfo{author}{\bibfnamefont{M.}~\bibnamefont{Sayad}} \bibnamefont{and}
  \bibinfo{author}{\bibfnamefont{M.}~\bibnamefont{Potthoff}},
  \bibinfo{journal}{New J. Phys.} \textbf{\bibinfo{volume}{17}},
  \bibinfo{pages}{113058} (\bibinfo{year}{2015}).

\bibitem[{\citenamefont{Sayad et~al.}(2016)\citenamefont{Sayad, Rausch, and
  Potthoff}}]{SRP16a}
\bibinfo{author}{\bibfnamefont{M.}~\bibnamefont{Sayad}},
  \bibinfo{author}{\bibfnamefont{R.}~\bibnamefont{Rausch}}, \bibnamefont{and}
  \bibinfo{author}{\bibfnamefont{M.}~\bibnamefont{Potthoff}},
  \bibinfo{journal}{Phys. Rev. Lett.} \textbf{\bibinfo{volume}{117}},
  \bibinfo{pages}{127201} (\bibinfo{year}{2016}).

\bibitem[{\citenamefont{Chern et~al.}(2018)\citenamefont{Chern, Barros, Wang,
  Suwa, and Batista}}]{CBW+18}
\bibinfo{author}{\bibfnamefont{G.-W.} \bibnamefont{Chern}},
  \bibinfo{author}{\bibfnamefont{K.}~\bibnamefont{Barros}},
  \bibinfo{author}{\bibfnamefont{Z.}~\bibnamefont{Wang}},
  \bibinfo{author}{\bibfnamefont{H.}~\bibnamefont{Suwa}}, \bibnamefont{and}
  \bibinfo{author}{\bibfnamefont{C.~D.} \bibnamefont{Batista}},
  \bibinfo{journal}{Phys. Rev. B} \textbf{\bibinfo{volume}{97}},
  \bibinfo{pages}{035120} (\bibinfo{year}{2018}).

\bibitem[{\citenamefont{Bajpai and Nikolic}(2019)}]{BN19}
\bibinfo{author}{\bibfnamefont{U.}~\bibnamefont{Bajpai}} \bibnamefont{and}
  \bibinfo{author}{\bibfnamefont{B.~K.} \bibnamefont{Nikolic}},
  \bibinfo{journal}{Phys. Rev. B} \textbf{\bibinfo{volume}{99}},
  \bibinfo{pages}{134409} (\bibinfo{year}{2019}).

\bibitem[{\citenamefont{Elze}(2012)}]{Elz12}
\bibinfo{author}{\bibfnamefont{H.}~\bibnamefont{Elze}}, \bibinfo{journal}{Phys.
  Rev. A} \textbf{\bibinfo{volume}{85}}, \bibinfo{pages}{052109}
  (\bibinfo{year}{2012}).

\bibitem[{\citenamefont{Berry}(1984)}]{Ber84}
\bibinfo{author}{\bibfnamefont{M.~V.} \bibnamefont{Berry}},
  \bibinfo{journal}{Proc. R. Soc. London A} \textbf{\bibinfo{volume}{392}},
  \bibinfo{pages}{45} (\bibinfo{year}{1984}).

\bibitem[{\citenamefont{Kuratsuji and Iida}(1985)}]{KI85}
\bibinfo{author}{\bibfnamefont{H.}~\bibnamefont{Kuratsuji}} \bibnamefont{and}
  \bibinfo{author}{\bibfnamefont{S.}~\bibnamefont{Iida}},
  \bibinfo{journal}{Prog. Theor. Phys.} \textbf{\bibinfo{volume}{74}},
  \bibinfo{pages}{439} (\bibinfo{year}{1985}).

\bibitem[{\citenamefont{Moody et~al.}(1986)\citenamefont{Moody, Shapere, and
  Wilczek}}]{MSW86}
\bibinfo{author}{\bibfnamefont{J.}~\bibnamefont{Moody}},
  \bibinfo{author}{\bibfnamefont{A.}~\bibnamefont{Shapere}}, \bibnamefont{and}
  \bibinfo{author}{\bibfnamefont{F.}~\bibnamefont{Wilczek}},
  \bibinfo{journal}{Phys. Rev. Lett.} \textbf{\bibinfo{volume}{56}},
  \bibinfo{pages}{893} (\bibinfo{year}{1986}).

\bibitem[{\citenamefont{Zygelman}(1987)}]{Zyg87}
\bibinfo{author}{\bibfnamefont{B.}~\bibnamefont{Zygelman}},
  \bibinfo{journal}{Phys. Lett. A} \textbf{\bibinfo{volume}{125}},
  \bibinfo{pages}{476} (\bibinfo{year}{1987}).

\bibitem[{\citenamefont{Stahl and Potthoff}(2017)}]{SP17}
\bibinfo{author}{\bibfnamefont{C.}~\bibnamefont{Stahl}} \bibnamefont{and}
  \bibinfo{author}{\bibfnamefont{M.}~\bibnamefont{Potthoff}},
  \bibinfo{journal}{Phys. Rev. Lett.} \textbf{\bibinfo{volume}{119}},
  \bibinfo{pages}{227203} (\bibinfo{year}{2017}).

\bibitem[{\citenamefont{Simon}(1983)}]{Sim83}
\bibinfo{author}{\bibfnamefont{B.}~\bibnamefont{Simon}},
  \bibinfo{journal}{Phys. Rev. Lett.} \textbf{\bibinfo{volume}{51}},
  \bibinfo{pages}{2167} (\bibinfo{year}{1983}).

\bibitem[{\citenamefont{Wilczek and Zee}(1984)}]{WZ84}
\bibinfo{author}{\bibfnamefont{F.}~\bibnamefont{Wilczek}} \bibnamefont{and}
  \bibinfo{author}{\bibfnamefont{A.}~\bibnamefont{Zee}},
  \bibinfo{journal}{Phys. Rev. Lett.} \textbf{\bibinfo{volume}{52}},
  \bibinfo{pages}{2111} (\bibinfo{year}{1984}).

\bibitem[{\citenamefont{Bohm et~al.}(2003)\citenamefont{Bohm, Mostafazadeh,
  Koizumi, Niu, and Zwanziger}}]{BMK+03}
\bibinfo{author}{\bibfnamefont{A.}~\bibnamefont{Bohm}},
  \bibinfo{author}{\bibfnamefont{A.}~\bibnamefont{Mostafazadeh}},
  \bibinfo{author}{\bibfnamefont{H.}~\bibnamefont{Koizumi}},
  \bibinfo{author}{\bibfnamefont{Q.}~\bibnamefont{Niu}}, \bibnamefont{and}
  \bibinfo{author}{\bibfnamefont{J.}~\bibnamefont{Zwanziger}},
  \emph{\bibinfo{title}{The Geometric Phase in Quantum Systems}}
  (\bibinfo{publisher}{Springer}, \bibinfo{address}{Berlin},
  \bibinfo{year}{2003}).

\bibitem[{\citenamefont{Berry and Robbins}(1993)}]{BR93b}
\bibinfo{author}{\bibfnamefont{M.}~\bibnamefont{Berry}} \bibnamefont{and}
  \bibinfo{author}{\bibfnamefont{J.}~\bibnamefont{Robbins}},
  \bibinfo{journal}{Proc. R. Soc. London A} \textbf{\bibinfo{volume}{442}},
  \bibinfo{pages}{659} (\bibinfo{year}{1993}).

\bibitem[{\citenamefont{Campisi et~al.}(2012)\citenamefont{Campisi, Denisov,
  and H\"anggi}}]{CDH12}
\bibinfo{author}{\bibfnamefont{M.}~\bibnamefont{Campisi}},
  \bibinfo{author}{\bibfnamefont{S.}~\bibnamefont{Denisov}}, \bibnamefont{and}
  \bibinfo{author}{\bibfnamefont{P.}~\bibnamefont{H\"anggi}},
  \bibinfo{journal}{Phys. Rev. A} \textbf{\bibinfo{volume}{86}},
  \bibinfo{pages}{032114} (\bibinfo{year}{2012}).

\bibitem[{\citenamefont{Skubic et~al.}(2008)\citenamefont{Skubic, Hellsvik,
  Nordstr\"om, and Eriksson}}]{SHNE08}
\bibinfo{author}{\bibfnamefont{B.}~\bibnamefont{Skubic}},
  \bibinfo{author}{\bibfnamefont{J.}~\bibnamefont{Hellsvik}},
  \bibinfo{author}{\bibfnamefont{L.}~\bibnamefont{Nordstr\"om}},
  \bibnamefont{and} \bibinfo{author}{\bibfnamefont{O.}~\bibnamefont{Eriksson}},
  \bibinfo{journal}{J. Phys.: Condens. Matter} \textbf{\bibinfo{volume}{20}},
  \bibinfo{pages}{315203} (\bibinfo{year}{2008}).

\bibitem[{llg()}]{llg}
\bibinfo{note}{L.~D. Landau and E.~M. Lifshitz, Physik. Zeits. Sowjetunion
  \textbf{8},153 (1935); T. Gilbert, Phys. Rev. \textbf{100}, 1243 (1955); T.
  Gilbert, Magnetics, IEEE Transactions on \textbf{40}, 3443 (2004).}

\bibitem[{\citenamefont{Michel and Potthoff}(2022)}]{MP22}
\bibinfo{author}{\bibfnamefont{S.}~\bibnamefont{Michel}} \bibnamefont{and}
  \bibinfo{author}{\bibfnamefont{M.}~\bibnamefont{Potthoff}},
  \bibinfo{journal}{Phys. Rev. B} \textbf{\bibinfo{volume}{106}},
  \bibinfo{pages}{235423} (\bibinfo{year}{2022}).

\bibitem[{\citenamefont{Ihm}(1991)}]{Ihm91}
\bibinfo{author}{\bibfnamefont{J.}~\bibnamefont{Ihm}}, \bibinfo{journal}{Phys.
  Rev. Lett.} \textbf{\bibinfo{volume}{67}}, \bibinfo{pages}{251}
  (\bibinfo{year}{1991}).

\bibitem[{\citenamefont{Bajpai and Nikoli\ifmmode~\acute{c}\else
  \'{c}\fi{}}(2020)}]{BN20}
\bibinfo{author}{\bibfnamefont{U.}~\bibnamefont{Bajpai}} \bibnamefont{and}
  \bibinfo{author}{\bibfnamefont{B.~K.}
  \bibnamefont{Nikoli\ifmmode~\acute{c}\else \'{c}\fi{}}},
  \bibinfo{journal}{Phys. Rev. Lett.} \textbf{\bibinfo{volume}{125}},
  \bibinfo{pages}{187202} (\bibinfo{year}{2020}).

\bibitem[{\citenamefont{Lenzing et~al.}(2022)\citenamefont{Lenzing,
  Lichtenstein, and Potthoff}}]{LLP22}
\bibinfo{author}{\bibfnamefont{N.}~\bibnamefont{Lenzing}},
  \bibinfo{author}{\bibfnamefont{A.~I.} \bibnamefont{Lichtenstein}},
  \bibnamefont{and} \bibinfo{author}{\bibfnamefont{M.}~\bibnamefont{Potthoff}},
  \bibinfo{journal}{Phys. Rev. B} \textbf{\bibinfo{volume}{106}},
  \bibinfo{pages}{094433} (\bibinfo{year}{2022}).

\bibitem[{\citenamefont{Elbracht et~al.}(2020)\citenamefont{Elbracht, Michel,
  and Potthoff}}]{EMP20}
\bibinfo{author}{\bibfnamefont{M.}~\bibnamefont{Elbracht}},
  \bibinfo{author}{\bibfnamefont{S.}~\bibnamefont{Michel}}, \bibnamefont{and}
  \bibinfo{author}{\bibfnamefont{M.}~\bibnamefont{Potthoff}},
  \bibinfo{journal}{Phys. Rev. Lett.} \textbf{\bibinfo{volume}{124}},
  \bibinfo{pages}{197202} (\bibinfo{year}{2020}).

\bibitem[{\citenamefont{Michel and Potthoff}(2021)}]{MP21}
\bibinfo{author}{\bibfnamefont{S.}~\bibnamefont{Michel}} \bibnamefont{and}
  \bibinfo{author}{\bibfnamefont{M.}~\bibnamefont{Potthoff}},
  \bibinfo{journal}{Phys. Rev. B} \textbf{\bibinfo{volume}{103}},
  \bibinfo{pages}{024449} (\bibinfo{year}{2021}).

\bibitem[{\citenamefont{Hannay}(1985)}]{Han85}
\bibinfo{author}{\bibfnamefont{J.~H.} \bibnamefont{Hannay}},
  \bibinfo{journal}{J. Phys. A} \textbf{\bibinfo{volume}{18}},
  \bibinfo{pages}{221} (\bibinfo{year}{1985}).

\bibitem[{\citenamefont{Haldane}(1988)}]{Hal88}
\bibinfo{author}{\bibfnamefont{F.~D.~M.} \bibnamefont{Haldane}},
  \bibinfo{journal}{Phys. Rev. Lett.} \textbf{\bibinfo{volume}{61}},
  \bibinfo{pages}{2015} (\bibinfo{year}{1988}).

\bibitem[{\citenamefont{Bernevig}(2013)}]{Ber13}
\bibinfo{author}{\bibfnamefont{B.~A.} \bibnamefont{Bernevig}},
  \emph{\bibinfo{title}{Topological insulators and topological
  superconductors}} (\bibinfo{publisher}{Princeton University Press},
  \bibinfo{address}{Princeton}, \bibinfo{year}{2013}).

\bibitem[{\citenamefont{Berry and Robbins}(1992)}]{BR92}
\bibinfo{author}{\bibfnamefont{M.}~\bibnamefont{Berry}} \bibnamefont{and}
  \bibinfo{author}{\bibfnamefont{J.}~\bibnamefont{Robbins}},
  \bibinfo{journal}{Proc. R. Soc. London A} \textbf{\bibinfo{volume}{436}},
  \bibinfo{pages}{631} (\bibinfo{year}{1992}).

\bibitem[{\citenamefont{Lenzing and Potthoff}(2023)}]{LP23}
\bibinfo{author}{\bibfnamefont{N.}~\bibnamefont{Lenzing}} \bibnamefont{and}
  \bibinfo{author}{\bibfnamefont{M.}~\bibnamefont{Potthoff}},
  \bibinfo{journal}{unpublished}  (\bibinfo{year}{2023}).

\bibitem[{\citenamefont{Gebhard}(1997)}]{Geb97}
\bibinfo{author}{\bibfnamefont{F.}~\bibnamefont{Gebhard}},
  \emph{\bibinfo{title}{The Mott Metal-Insulator Transition}}
  (\bibinfo{publisher}{Springer}, \bibinfo{address}{Berlin},
  \bibinfo{year}{1997}).

\bibitem[{\citenamefont{Essler et~al.}(2005)\citenamefont{Essler, Frahm,
  G\"ohmann, Kl\"umper, and Korepin}}]{EFG+05}
\bibinfo{author}{\bibfnamefont{F.~H.~L.} \bibnamefont{Essler}},
  \bibinfo{author}{\bibfnamefont{H.}~\bibnamefont{Frahm}},
  \bibinfo{author}{\bibfnamefont{F.}~\bibnamefont{G\"ohmann}},
  \bibinfo{author}{\bibfnamefont{A.}~\bibnamefont{Kl\"umper}},
  \bibnamefont{and} \bibinfo{author}{\bibfnamefont{V.}~\bibnamefont{Korepin}},
  \emph{\bibinfo{title}{The One-Dimensional Hubbard Model}}
  (\bibinfo{publisher}{Cambridge University Press},
  \bibinfo{address}{Cambridge}, \bibinfo{year}{2005}).

\bibitem[{\citenamefont{Anderson}(1950)}]{And50}
\bibinfo{author}{\bibfnamefont{P.~W.} \bibnamefont{Anderson}},
  \bibinfo{journal}{Phys. Rev.} \textbf{\bibinfo{volume}{79}},
  \bibinfo{pages}{350} (\bibinfo{year}{1950}).

\bibitem[{\citenamefont{Schrieffer et~al.}(1989)\citenamefont{Schrieffer, Wen,
  and Zhang}}]{SWZ89}
\bibinfo{author}{\bibfnamefont{J.~R.} \bibnamefont{Schrieffer}},
  \bibinfo{author}{\bibfnamefont{X.~G.} \bibnamefont{Wen}}, \bibnamefont{and}
  \bibinfo{author}{\bibfnamefont{S.~C.} \bibnamefont{Zhang}},
  \bibinfo{journal}{Phys. Rev. B} \textbf{\bibinfo{volume}{39}},
  \bibinfo{pages}{11663} (\bibinfo{year}{1989}).

\bibitem[{\citenamefont{Manousakis}(1991)}]{Man91}
\bibinfo{author}{\bibfnamefont{E.}~\bibnamefont{Manousakis}},
  \bibinfo{journal}{Rev. Mod. Phys.} \textbf{\bibinfo{volume}{63}},
  \bibinfo{pages}{1} (\bibinfo{year}{1991}).

\bibitem[{\citenamefont{Auerbach}(1994)}]{Aue94}
\bibinfo{author}{\bibfnamefont{A.}~\bibnamefont{Auerbach}},
  \emph{\bibinfo{title}{Interacting electrons and quantum magnetism}}
  (\bibinfo{publisher}{Springer}, \bibinfo{address}{New York},
  \bibinfo{year}{1994}).

\bibitem[{\citenamefont{Sch\"afer et~al.}(2015)\citenamefont{Sch\"afer, Geles,
  Rost, Rohringer, Arrigoni, Held, Bl\"umer, Aichhorn, and Toschi}}]{SGR+15}
\bibinfo{author}{\bibfnamefont{T.}~\bibnamefont{Sch\"afer}},
  \bibinfo{author}{\bibfnamefont{F.}~\bibnamefont{Geles}},
  \bibinfo{author}{\bibfnamefont{D.}~\bibnamefont{Rost}},
  \bibinfo{author}{\bibfnamefont{G.}~\bibnamefont{Rohringer}},
  \bibinfo{author}{\bibfnamefont{E.}~\bibnamefont{Arrigoni}},
  \bibinfo{author}{\bibfnamefont{K.}~\bibnamefont{Held}},
  \bibinfo{author}{\bibfnamefont{N.}~\bibnamefont{Bl\"umer}},
  \bibinfo{author}{\bibfnamefont{M.}~\bibnamefont{Aichhorn}}, \bibnamefont{and}
  \bibinfo{author}{\bibfnamefont{A.}~\bibnamefont{Toschi}},
  \bibinfo{journal}{Phys. Rev. B} \textbf{\bibinfo{volume}{91}},
  \bibinfo{pages}{125109} (\bibinfo{year}{2015}).

\bibitem[{\citenamefont{Moriya}(1985)}]{Mor85}
\bibinfo{author}{\bibfnamefont{T.}~\bibnamefont{Moriya}},
  \emph{\bibinfo{title}{Spin Fluctuations in Itinerant Electron Magnetism}},
  vol.~\bibinfo{volume}{56} of \emph{\bibinfo{series}{Springer Series in
  Solid-State Sciences}} (\bibinfo{publisher}{Springer},
  \bibinfo{address}{Berlin}, \bibinfo{year}{1985}).

\bibitem[{\citenamefont{Rowe et~al.}(2012)\citenamefont{Rowe, Knolle, Eremin,
  and Hirschfeld}}]{RKEH12}
\bibinfo{author}{\bibfnamefont{W.}~\bibnamefont{Rowe}},
  \bibinfo{author}{\bibfnamefont{J.}~\bibnamefont{Knolle}},
  \bibinfo{author}{\bibfnamefont{I.}~\bibnamefont{Eremin}}, \bibnamefont{and}
  \bibinfo{author}{\bibfnamefont{P.~J.} \bibnamefont{Hirschfeld}},
  \bibinfo{journal}{Phys. Rev. B} \textbf{\bibinfo{volume}{86}},
  \bibinfo{pages}{134513} (\bibinfo{year}{2012}).

\bibitem[{\citenamefont{Del~Re and Toschi}(2021)}]{DT21}
\bibinfo{author}{\bibfnamefont{L.}~\bibnamefont{Del~Re}} \bibnamefont{and}
  \bibinfo{author}{\bibfnamefont{A.}~\bibnamefont{Toschi}},
  \bibinfo{journal}{Phys. Rev. B} \textbf{\bibinfo{volume}{104}},
  \bibinfo{pages}{085120} (\bibinfo{year}{2021}).

\bibitem[{\citenamefont{Luttinger and Ward}(1960)}]{LW60}
\bibinfo{author}{\bibfnamefont{J.~M.} \bibnamefont{Luttinger}}
  \bibnamefont{and} \bibinfo{author}{\bibfnamefont{J.~C.} \bibnamefont{Ward}},
  \bibinfo{journal}{Phys. Rev.} \textbf{\bibinfo{volume}{118}},
  \bibinfo{pages}{1417} (\bibinfo{year}{1960}).

\bibitem[{\citenamefont{Rohringer et~al.}(2018)\citenamefont{Rohringer,
  Hafermann, Toschi, Katanin, Antipov, Katsnelson, Lichtenstein, Rubtsov, and
  Held}}]{RHT+18}
\bibinfo{author}{\bibfnamefont{G.}~\bibnamefont{Rohringer}},
  \bibinfo{author}{\bibfnamefont{H.}~\bibnamefont{Hafermann}},
  \bibinfo{author}{\bibfnamefont{A.}~\bibnamefont{Toschi}},
  \bibinfo{author}{\bibfnamefont{A.~A.} \bibnamefont{Katanin}},
  \bibinfo{author}{\bibfnamefont{A.~E.} \bibnamefont{Antipov}},
  \bibinfo{author}{\bibfnamefont{M.~I.} \bibnamefont{Katsnelson}},
  \bibinfo{author}{\bibfnamefont{A.~I.} \bibnamefont{Lichtenstein}},
  \bibinfo{author}{\bibfnamefont{A.~N.} \bibnamefont{Rubtsov}},
  \bibnamefont{and} \bibinfo{author}{\bibfnamefont{K.}~\bibnamefont{Held}},
  \bibinfo{journal}{Rev. Mod. Phys.} \textbf{\bibinfo{volume}{90}},
  \bibinfo{pages}{025003} (\bibinfo{year}{2018}).

\bibitem[{\citenamefont{Chakravarty et~al.}(1989)\citenamefont{Chakravarty,
  Halperin, and Nelson}}]{SHN89}
\bibinfo{author}{\bibfnamefont{S.}~\bibnamefont{Chakravarty}},
  \bibinfo{author}{\bibfnamefont{B.~I.} \bibnamefont{Halperin}},
  \bibnamefont{and} \bibinfo{author}{\bibfnamefont{D.~R.}
  \bibnamefont{Nelson}}, \bibinfo{journal}{Phys. Rev. B}
  \textbf{\bibinfo{volume}{39}}, \bibinfo{pages}{2344} (\bibinfo{year}{1989}).

\bibitem[{\citenamefont{Sandvik}(1997)}]{San97}
\bibinfo{author}{\bibfnamefont{A.~W.} \bibnamefont{Sandvik}},
  \bibinfo{journal}{Phys. Rev. B} \textbf{\bibinfo{volume}{56}},
  \bibinfo{pages}{11678} (\bibinfo{year}{1997}).

\bibitem[{\citenamefont{Anderson}(1952)}]{And52}
\bibinfo{author}{\bibfnamefont{P.~W.} \bibnamefont{Anderson}},
  \bibinfo{journal}{Phys. Rev.} \textbf{\bibinfo{volume}{86}},
  \bibinfo{pages}{694} (\bibinfo{year}{1952}).

\bibitem[{\citenamefont{Holstein and Primakoff}(1940)}]{HP40}
\bibinfo{author}{\bibfnamefont{J.}~\bibnamefont{Holstein}} \bibnamefont{and}
  \bibinfo{author}{\bibfnamefont{N.}~\bibnamefont{Primakoff}},
  \bibinfo{journal}{Phys. Rev.} \textbf{\bibinfo{volume}{58}},
  \bibinfo{pages}{1908} (\bibinfo{year}{1940}).

\bibitem[{\citenamefont{Hamer et~al.}(1992)\citenamefont{Hamer, Weihong, and
  Arndt}}]{HWA92}
\bibinfo{author}{\bibfnamefont{C.~J.} \bibnamefont{Hamer}},
  \bibinfo{author}{\bibfnamefont{Z.}~\bibnamefont{Weihong}}, \bibnamefont{and}
  \bibinfo{author}{\bibfnamefont{P.}~\bibnamefont{Arndt}},
  \bibinfo{journal}{Phys. Rev. B} \textbf{\bibinfo{volume}{46}},
  \bibinfo{pages}{6276} (\bibinfo{year}{1992}).

\bibitem[{\citenamefont{Chernyshev and Maksimov}(2016)}]{CM16}
\bibinfo{author}{\bibfnamefont{A.~L.} \bibnamefont{Chernyshev}}
  \bibnamefont{and} \bibinfo{author}{\bibfnamefont{P.~A.}
  \bibnamefont{Maksimov}}, \bibinfo{journal}{Phys. Rev. Lett.}
  \textbf{\bibinfo{volume}{117}}, \bibinfo{pages}{187203}
  (\bibinfo{year}{2016}).

\bibitem[{\citenamefont{McClarty et~al.}(2018)\citenamefont{McClarty, Dong,
  Gohlke, Rau, Pollmann, Moessner, and Penc}}]{MDG+18}
\bibinfo{author}{\bibfnamefont{P.~A.} \bibnamefont{McClarty}},
  \bibinfo{author}{\bibfnamefont{X.-Y.} \bibnamefont{Dong}},
  \bibinfo{author}{\bibfnamefont{M.}~\bibnamefont{Gohlke}},
  \bibinfo{author}{\bibfnamefont{J.~G.} \bibnamefont{Rau}},
  \bibinfo{author}{\bibfnamefont{F.}~\bibnamefont{Pollmann}},
  \bibinfo{author}{\bibfnamefont{R.}~\bibnamefont{Moessner}}, \bibnamefont{and}
  \bibinfo{author}{\bibfnamefont{K.}~\bibnamefont{Penc}},
  \bibinfo{journal}{Phys. Rev. B} \textbf{\bibinfo{volume}{98}},
  \bibinfo{pages}{060404} (\bibinfo{year}{2018}).

\bibitem[{\citenamefont{McClarty}(2022)}]{McC22}
\bibinfo{author}{\bibfnamefont{P.}~\bibnamefont{McClarty}},
  \bibinfo{journal}{Annu. Rev. Condens. Matter Phys.}
  \textbf{\bibinfo{volume}{13}}, \bibinfo{pages}{171} (\bibinfo{year}{2022}).

\bibitem[{SM()}]{SM}
\bibinfo{note}{See Supplemental Material at {\em URL} for various technical details and supplemental results.}

\bibitem[{\citenamefont{Goldstone}(1961)}]{Gol61}
\bibinfo{author}{\bibfnamefont{J.}~\bibnamefont{Goldstone}},
  \bibinfo{journal}{Nuovo Cim.} \textbf{\bibinfo{volume}{19}},
  \bibinfo{pages}{154} (\bibinfo{year}{1961}).

\bibitem[{\citenamefont{Nambu and Jona-Lasinio}(1961)}]{YJ61}
\bibinfo{author}{\bibfnamefont{Y.}~\bibnamefont{Nambu}} \bibnamefont{and}
  \bibinfo{author}{\bibfnamefont{G.}~\bibnamefont{Jona-Lasinio}},
  \bibinfo{journal}{Phys. Rev.} \textbf{\bibinfo{volume}{122}},
  \bibinfo{pages}{345} (\bibinfo{year}{1961}).

\bibitem[{\citenamefont{Brauner}(2010)}]{Bra10}
\bibinfo{author}{\bibfnamefont{T.}~\bibnamefont{Brauner}},
  \bibinfo{journal}{Symmetry} \textbf{\bibinfo{volume}{2}},
  \bibinfo{pages}{609} (\bibinfo{year}{2010}).

\bibitem[{\citenamefont{Mermin and Wagner}(1966)}]{MW66}
\bibinfo{author}{\bibfnamefont{N.~D.} \bibnamefont{Mermin}} \bibnamefont{and}
  \bibinfo{author}{\bibfnamefont{H.}~\bibnamefont{Wagner}},
  \bibinfo{journal}{Phys. Rev. Lett.} \textbf{\bibinfo{volume}{17}},
  \bibinfo{pages}{1133} (\bibinfo{year}{1966}).

\bibitem[{\citenamefont{Metzner and Vollhardt}(1989)}]{MV89}
\bibinfo{author}{\bibfnamefont{W.}~\bibnamefont{Metzner}} \bibnamefont{and}
  \bibinfo{author}{\bibfnamefont{D.}~\bibnamefont{Vollhardt}},
  \bibinfo{journal}{Phys. Rev. Lett.} \textbf{\bibinfo{volume}{62}},
  \bibinfo{pages}{324} (\bibinfo{year}{1989}).

\bibitem[{\citenamefont{M\"uller-Hartmann}(1989)}]{MH89b}
\bibinfo{author}{\bibfnamefont{E.}~\bibnamefont{M\"uller-Hartmann}},
  \bibinfo{journal}{Z. Phys. B} \textbf{\bibinfo{volume}{74}},
  \bibinfo{pages}{507} (\bibinfo{year}{1989}).

\bibitem[{\citenamefont{Oguchi}(1960)}]{Ogu60}
\bibinfo{author}{\bibfnamefont{T.}~\bibnamefont{Oguchi}},
  \bibinfo{journal}{Phys. Rev.} \textbf{\bibinfo{volume}{117}},
  \bibinfo{pages}{117} (\bibinfo{year}{1960}).

\bibitem[{\citenamefont{Liu}(1990)}]{BG90}
\bibinfo{author}{\bibfnamefont{B.-G.} \bibnamefont{Liu}},
  \bibinfo{journal}{Phys. Rev. B} \textbf{\bibinfo{volume}{41}},
  \bibinfo{pages}{9563} (\bibinfo{year}{1990}).

\bibitem[{\citenamefont{Majlis et~al.}(1992)\citenamefont{Majlis, Selzer, and
  Strinati}}]{MSS92}
\bibinfo{author}{\bibfnamefont{N.}~\bibnamefont{Majlis}},
  \bibinfo{author}{\bibfnamefont{S.}~\bibnamefont{Selzer}}, \bibnamefont{and}
  \bibinfo{author}{\bibfnamefont{G.~C.} \bibnamefont{Strinati}},
  \bibinfo{journal}{Phys. Rev. B} \textbf{\bibinfo{volume}{45}},
  \bibinfo{pages}{7872} (\bibinfo{year}{1992}).

\bibitem[{\citenamefont{Majlis et~al.}(1993)\citenamefont{Majlis, Selzer, and
  Strinati}}]{MSS93}
\bibinfo{author}{\bibfnamefont{N.}~\bibnamefont{Majlis}},
  \bibinfo{author}{\bibfnamefont{S.}~\bibnamefont{Selzer}}, \bibnamefont{and}
  \bibinfo{author}{\bibfnamefont{G.~C.} \bibnamefont{Strinati}},
  \bibinfo{journal}{Phys. Rev. B} \textbf{\bibinfo{volume}{48}},
  \bibinfo{pages}{957} (\bibinfo{year}{1993}).

\end{thebibliography}
\end{document}